\documentclass[
reprint,
superscriptaddress,
 amsmath,amssymb,
 aps,
prl,
]{revtex4-1}

\usepackage{graphicx}
\usepackage{dcolumn}
\usepackage{bm}
\usepackage{makecell}
\usepackage{relsize}
\usepackage{array}
\usepackage{mathtools}
\usepackage{url}
\usepackage{hyperref}

\renewcommand{\Re}{\operatorname{Re}}
\renewcommand{\Im}{\operatorname{Im}}

\begin{document}

\title{Floquet engineering of nonreciprocal light-induced dipolar interactions}

\author{Livia Egyed}
\thanks{These authors contributed equally.}
\affiliation{Vienna Center for Quantum Science and Technology (VCQ), Atominstitut, TU Wien, Stadionallee 2, A-1020 Vienna, Austria}

\author{Murad Abuzarli}
\thanks{These authors contributed equally.}
\affiliation{Vienna Center for Quantum Science and Technology (VCQ), Faculty of Physics, University of Vienna, Boltzmanngasse 5, A-1090 Vienna, Austria}

\author{Manuel Reisenbauer}
\affiliation{Vienna Center for Quantum Science and Technology (VCQ), Atominstitut, TU Wien, Stadionallee 2, A-1020 Vienna, Austria}
\affiliation{Vienna Center for Quantum Science and Technology (VCQ), Faculty of Physics, University of Vienna, Boltzmanngasse 5, A-1090 Vienna, Austria}

\author{Iurie Coroli}
\affiliation{Vienna Center for Quantum Science and Technology (VCQ), Faculty of Physics, University of Vienna, Boltzmanngasse 5, A-1090 Vienna, Austria}

\author{Benjamin A. Stickler}
\affiliation{Institute for Complex Quantum Systems, Ulm University, Albert-Einstein-Allee 11, D-89069 Ulm, Germany}

\author{Uro\v{s} Deli\'{c}}
\email{uros.delic@tuwien.ac.at}
\affiliation{Vienna Center for Quantum Science and Technology (VCQ), Atominstitut, TU Wien, Stadionallee 2, A-1020 Vienna, Austria}

\date{\today}

\begin{abstract}

Tweezer arrays of polarizable objects are a promising platform for assembling quantum matter and building next-generation quantum sensors. Light-induced dipolar interactions have emerged as a method to couple their motion, thereby establishing a new paradigm for controlling collective mechanical degrees of freedom. Here, we extend these into the regime of Floquet-driven interactions, combined with the intrinsic nonreciprocity of optical forces. We demonstrate beamsplitter, single-, and two-mode squeezing operations, as well as signatures of a negative-mass-like oscillator arising from the nonreciprocity. Moreover, we show that a programmable combination of these operations enables continuous tuning of complex eigenfrequencies. These results establish a toolbox of quantum operations of nonreciprocal interactions that are essential for investigating non-Hermitian many-body physics and collective quantum optomechanics.

\end{abstract}

\maketitle

Controllable interactions between quantum systems are the backbone of state-of-the-art experiments in quantum metrology \cite{ProgressQuantumMetrology}, quantum sensing \cite{YeZollerEssay,ColomboEntanglementClocks}, and quantum many-body physics \cite{BlochRMPManyBody}. On atomic and molecular platforms, success has been driven by dipolar, Rydberg, and Coulomb interactions that enable spin exchange and correlations between nearby atoms or ions \cite{DefenuRMP, ChomazReviewDipolar, SaffmanReviewRydberg, MonroeReviewIons}. On the other hand, light-mediated interactions enable interfaces between systems over large distances. Therefore, they have been used in optomechanics, enabling the coupling of macroscopic mechanical oscillators via an optical cavity \cite{ShkarinHybridization, Fang2017, MathewEwold, GroeblacherArrays, Silanpää_2021, Kotler2021, ChegnizadehScience}, and in hybrid systems to couple oscillators to atoms \cite{TreutleinPRL, Karg2020, PolzikEntanglement}. When reservoir engineering is employed, e.g., by coupling the oscillator to the collective spin of an atomic cloud or applying multiple cavity drives, the system can exhibit negative-mass dynamics, thereby enabling back-action evasion and quantum non-demolition measurements \cite{Silanpää_2016, Silanpää_2021, MollerBackAction, PolzikEntanglement, Steele_2024, HuardStabilizedSqueezing}.

For all polarizable objects, from atoms \cite{Maximo2018, Maiwoger2022} to dielectric bodies \cite{Dholakia2010, RudolphBindingPRA, RudolphBindingPRL}, illumination gives rise to a long-range, light-induced dipolar interaction between their mechanical degrees of freedom, the so-called “optical binding”. In recent years, it has been thoroughly explored in tweezer arrays of nanoparticles, demonstrating coupling of their positions in free space \cite{Rieser2022, LiskaColdDamping, TongBinding} or via an optical cavity \cite{vijayan_cavity-mediated_2024, PontinCavityMediated}. A notable feature of this interaction is its inherent nonreciprocity, arising from the engineered nonuniform optical phase of the drive across the tweezer sites. For instance, anti-reciprocal forces, characterized by equal magnitudes and directions, give rise to the non-Hermitian predator-prey dynamics of the nonlinear motion of two nanoparticles \cite{Liska2024, Reisenbauer2024}. However, the position coupling is limited in its applicability to quantum protocols \cite{Cavoptomechanics, BarzanjehReview} and non-Hermitian quantum physics \cite{Metelmann2015, metelmann_non_reci_entanglement_2023, SlimKitaev}, as it lacks deterministic access to the building blocks of quantum optics -- beamsplitter and two-mode squeezing operations.

In this work, we finally demonstrate these operations by engineering the optical frequencies across tweezer sites, thereby inducing a phase-modulated, Floquet-driven optical interaction. We show coherent energy exchange and correlated motion as signatures of the beamsplitter and two-mode squeezing operations between oscillators, respectively. We reveal that anti-reciprocal coupling realizes a negative-mass-like oscillator in the absence of atomic or cavity resonances, albeit in the rotating frame of interaction. Finally, we show a range of interactions “in-between”, in which their nature changes from fully reciprocal to anti-reciprocal periodically over time, enabling continuous tuning of both the real and imaginary parts of the mode eigenfrequencies. Beyond providing a novel control toolbox for light-mediated interactions between trapped nanoparticles, the demonstrated operations also enable coherent control in hybrid systems of nanoparticles and atoms or ions \cite{NanoparticleIon}, or between atoms \cite{Ho_selforganization-atoms, Wang_Braggscattering}.

Light-induced dipolar interactions exploit the phase coherence between the trapping light that induces the dipole in a polarizable object and the object's dipole radiation. Radiated light from one dipole travels a distance $d$ to the second dipole, where it interferes with the local trapping field. As the radiated field acquires a traveling phase $kd$, where $k$ is the wavenumber, the total trapping field experienced by the second dipole varies with distance. If the first dipole moves in any direction, the field at the location of the second dipole changes, thereby inducing optical forces that couple the motion of the induced dipoles. Thus, dipole 1 with position $z_1$ experiences a force proportional to $g_{12} z_2$ and dipole 2 with position $z_2$ experiences a force proportional to $g_{21} z_1$, where $g_{12}$ and $g_{21}$ are directional coupling rates. If the two objects are illuminated with equal optical phases, the forces are reciprocal with $g_{12}\equiv g_{21}$. Introducing an optical phase difference $\Delta\phi$ between the drives breaks the mirror symmetry of the system, yielding nonreciprocal forces and coupling rates that can be tuned by distance and phases \cite{Rieser2022, RudolphBindingPRA, RudolphBindingPRL}.

\begin{figure}[t]
    \includegraphics[width=\linewidth]{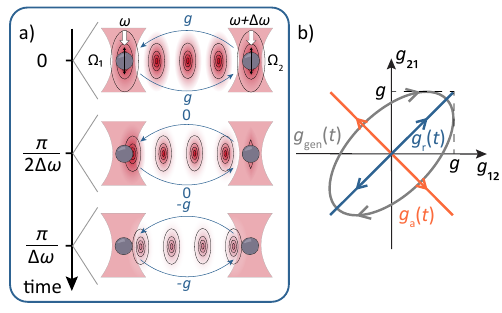}
    \caption{\textbf{Time-dependent light-induced dipolar interactions.}  a) Two particles with mechanical frequencies $\Omega_{1,2}$ in tweezers with different optical frequencies $\omega$ and $\omega+\Delta \omega$ interact via scattered light, which forms a phase-modulated standing wave and locally changes trapping potentials. The optical detuning yields time-dependent optical forces, so that the reciprocal coupling shown here changes from maximally positive to negative coupling rates after a time $\pi/\Delta \omega$. b) Time-modulated directional coupling rates $g_{12}$ and $g_{21}$ between the particles. Depending on the interparticle distance, the interaction can be reciprocal (blue), anti-reciprocal (orange), or their combination (gray). The absolute magnitude of coupling rates is bounded by $g$.}
    \label{fig1:interactions}
\end{figure}

Here, we go beyond the current state-of-the-art experiments by engineering the optical frequencies of the drives to differ by a detuning $\Delta \omega$. Consequently, the interference between the trapping and radiated fields oscillates at the same frequency, producing a modulated optical force for each object. This results in time-dependent directional coupling rates $g_{12}=g\cos(\Delta\omega t+\Delta\phi)$ and $g_{21}=g\cos(\Delta\omega t+\Delta\phi +2kd)$, where $g$ is the maximum coupling strength in the system \cite{SI}. The modulation creates a continuously rotating optical phase, thereby preventing the tuning of the interaction type via the relative optical phase. However, the nature of the interaction can still be controlled by the interparticle distance. As an example, setting $kd=n\pi$ ($n\in \mathbb{N}$) realizes reciprocal rates $g_{12} = g_{21} = \pm g \cos(\Delta \omega t+\Delta\phi)$ that periodically cycle between $g$ and $-g$ over a half-period of $\pi/\Delta\omega$ (Fig. \ref{fig1:interactions}a). For $kd=(2n+1) \pi/2$, the objects experience an anti-reciprocal interaction with $g_{12}=-g_{21}$ at all times, following the parametric function $g_{\text{a}}(t)$  (Fig. \ref{fig1:interactions}b). Finally, any intermediate distance $d$ yields coupling rates that are a programmable combination $g_\text{gen}(t)$ of the previous two cases.  Although similar in concept to the amplitude modulation used to nonreciprocally couple mechanical modes in cavity optomechanics \cite{Fang2017, WanjuraQuadrature, MathewEwold}, light-induced dipolar interactions between two objects are already nonreciprocal in the absence of a cavity, similar to the cascaded coupling in hybrid membrane-atom systems \cite{Karg2020, PolzikEntanglement}.

\begin{figure}[t]
    \centering
    \includegraphics[width=1\linewidth]{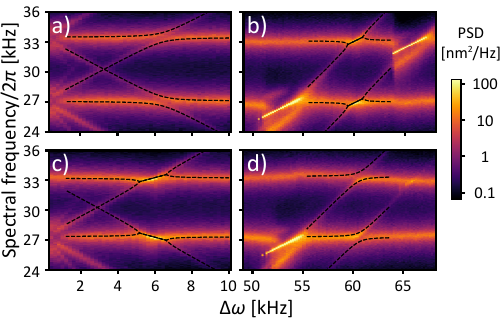}
    \caption{\textbf{Spectrograms of single- and two-mode operations.} Combined spectrograms of the motion $z_{1}$ and $z_2$ with frequencies $\Omega_1/2\pi\approx 27$ kHz and $\Omega_2/2\pi\approx 33$ kHz reveal interactions at different resonances: a) reciprocal beamsplitter at $\Delta\omega/2\pi\approx 6$ kHz, b) reciprocal two-mode squeezing at $\Delta\omega/2\pi\approx 60$ kHz, c) anti-reciprocal two-mode squeezing at $\Delta\omega/2\pi\approx 6$ kHz, d) anti-reciprocal beamsplitter interaction at $\Delta\omega/2\pi\approx 60$ kHz. Single-mode squeezing is observed in b) and d) at $\Delta\omega\approx 2\Omega_{1,2}$. The black lines are fits based on the theoretical dependence of the normal mode frequencies.}
    \label{fig2:scans}
\end{figure}

In the experiment, we trap two silica nanoparticles in distinct optical tweezers at a variable distance $d$ and at a pressure of $p = 1.2 \pm 0.1$ mbar, resulting in a mechanical damping rate of $\gamma/2\pi\approx 570\pm 40 ~\text{Hz}$. Their nominal radius of $r=105\pm 2$ nm is substantially smaller than the laser wavelength ($\lambda=1064$ nm), thus ensuring that they scatter light as almost ideal dipoles. We can independently control the optical detuning $\Delta \omega$ and the mechanical frequencies $\Omega_1$ and $\Omega_2$ \cite{SI}. 

\begin{figure*}[ht]
    \centering
    \includegraphics[width=\textwidth]{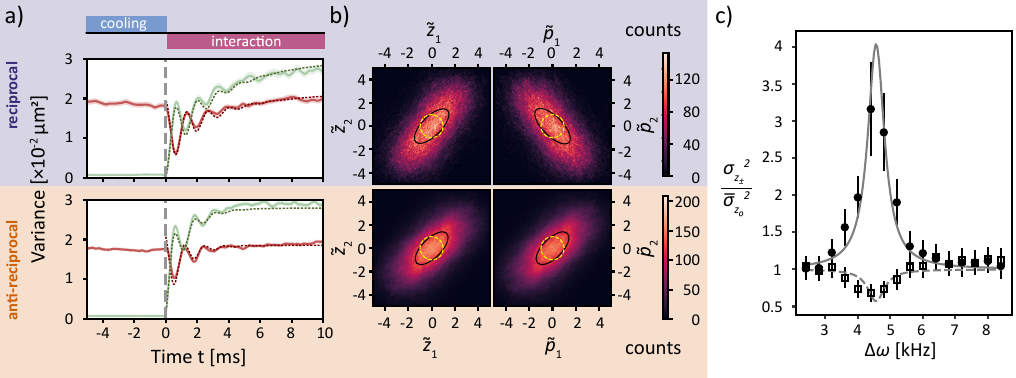}
    \caption{\textbf{Programmable two-mode operations for reciprocal and anti-reciprocal interactions.} a) Feedback cooling on the charged particle is switched off, while the beamsplitter interaction is switched on at time $t=0$ ms. Coherent energy exchange between reciprocally (top) and anti-reciprocally (bottom) coupled charged (green) and uncharged (red) particles measured at optical detunings $\Delta\omega\approx\Delta\Omega$ and $\Delta\omega\approx 2\bar\Omega$, respectively, and averaged over around $1000$ realizations. Dashed lines are theoretical fits. b) Stationary two-mode thermomechanical noise squashing between reciprocally coupled particles (top) shows correlations (anti-correlations) of scaled positions $\tilde{z}_{1,2}$ (momenta $\tilde{p}_{1,2}$). For anti-reciprocally coupled particles (bottom), both positions and momenta exhibit correlations. Solid black (dashed yellow) ellipses enclose regions of one standard deviation of resonant (non-resonant) two-mode squashed state. c) Normal-mode variances $\sigma^2_{\tilde{z}_-}$ (hollow squares) and $\sigma^2_{\tilde{z}_+}$ (solid circles) calibrated to thermal state variances $\bar{\sigma}^2_{z_0}$, fit with dashed and solid lines from the theory for anti-reciprocally coupled particles, show squeezing and anti-squeezing, respectively, as a function of the optical detuning $\Delta \omega$. The error bars are obtained from the fits in b) and are normalized to the area differences between the data points and the fit.} 
    \label{fig3:extremes}
\end{figure*}
The phase modulation at $\Delta \omega$ allows us to selectively enhance a specific interaction type while suppressing others via the rotating-wave approximation. For instance, setting $\Delta \omega$ in the vicinity of the mechanical detuning $\Delta \Omega=\Omega_2-\Omega_1$ mainly drives the following dynamics at an arbitrary distance $d$:
\begin{align}
    \left(\frac{\partial}{\partial t}+\frac{\gamma}{2}\right)
    \begin{pmatrix}
        b_1 \\
        b_2
    \end{pmatrix}=
    \frac{i}{2}
    \begin{pmatrix}
        \Delta\Omega-\Delta\omega
        &
        \tilde{g}\textrm{e}^{-ikd} 
        \\
        \tilde{g}^*\textrm{e}^{-ikd}
        &
        \Delta\omega-\Delta\Omega
    \end{pmatrix}
    \begin{pmatrix}
        b_1 \\
        b_2
    \end{pmatrix},
    \label{eq:RWAdynamics}
\end{align}
where $b_{j}$ are the phonon annihilation operators of displacements $z_j$ in the interaction frame, and $\tilde{g}=ge^{i\Delta\phi}$ \cite{SI}. For $kd=n\pi$, the interaction becomes reciprocal and realizes an effective beamsplitter interaction Hamiltonian $H_{\text{int}}=\hbar (\tilde{g}b_1^\dagger b_2 + \tilde{g}^*b_1b_2^\dagger)/2$, which can also be illustrated as a two-photon Raman scattering process. Namely, the exchange of photons from each tweezer via a virtual level provides an energy change of $\Delta \omega$, which compensates for the energy difference of $\Delta\Omega$ due to the addition and subtraction of a phonon on different oscillators. To observe the beamsplitter dynamics, we set the mechanical detuning $\Delta\Omega/2\pi\approx 6$ kHz while varying the optical detuning $\Delta \omega$ and monitoring the particle motion. As $\Delta\omega$ approaches $\Delta\Omega$, the time-dependent coupling makes the beamsplitter interaction resonant, resulting in an avoided crossing of the dressed modes in the joint spectrogram of particle positions (Fig. \ref{fig2:scans}a). We fit the mode eigenfrequencies in the vicinity of the avoided crossing to obtain the coupling rate $g/2\pi = 953  \pm 17$ Hz \cite{SI}.

A different resonance is obtained in the vicinity of the mechanical frequencies sum $2\bar{\Omega} =\Omega_1+\Omega_2$:
\begin{align}
    \left(\frac{\partial}{\partial t}+\frac{\gamma}{2}\right)
    \begin{pmatrix}
        b_1 \\
        b_2^\dagger
    \end{pmatrix}=
    \frac{i}{2}
    \begin{pmatrix}
        \Delta\omega-2\bar\Omega
        &
        \tilde{g}^*\textrm{e}^{ikd} 
        \\
        -\tilde{g}\textrm{e}^{ikd}
        &
        2\bar\Omega-\Delta\omega
    \end{pmatrix}
    \begin{pmatrix}
        b_1 \\
        b_2^\dagger
    \end{pmatrix}.
    \label{eq:RWAdynamics2}
\end{align}
For $kd=n\pi$, the reciprocal coupling yields a two-mode squeezing interaction Hamiltonian $H_{\text{int}}=\hbar (\tilde{g} b_1 b_2 + \tilde{g}^* b_1^\dagger b_2^\dagger)/2$ that creates or annihilates a pair of phonons \cite{MahboobTMS}. Consequently, the dressed mechanical modes become degenerate (Fig. \ref{fig2:scans}b), and fitting to the degeneracy region allows us to obtain a coupling rate of $g/2\pi= 806 \pm 24$ Hz. As $g>\gamma$, we observe increased oscillation amplitudes and nonlinear motion within the region \cite{Liska2024, Reisenbauer2024}. Finally, choosing $\Delta \omega$ to be approximately $2\Omega_1$ or $2\Omega_2$ yields a single-mode squeezing operation in a region around those frequencies. This effect is driven by the coupling but is applied to a single particle, in contrast to the standard case of modulating the intrinsic mechanical frequency \cite{Rugarsqueezing}.

Changing the distance to $kd=(2n+1)\pi/2$ realizes an anti-reciprocal interaction that cannot be described through a joint system Hamiltonian, as the dynamical matrices in Eqs. \eqref{eq:RWAdynamics} and \eqref{eq:RWAdynamics2} become non-Hermitian. However, we can formulate a pseudo-Hamiltonian by performing a non-canonical transformation $b_2\rightarrow \tilde{b}_2^\dagger$ on a single degree of freedom, which recovers an effective coupling between a positive-mass and a negative-mass oscillator \cite{SI}. Although our method differs from realizations of genuine negative-mass oscillators \cite{Karg2020, Silanpää_2021, MollerBackAction, Steele_2024}, the picture remains useful for understanding the resulting system dynamics. As a result, we now realize the two-mode squeezing operation at $\Delta \omega\approx \Delta \Omega$ (Fig. \ref{fig2:scans}c), in accordance with the picture of the swapped roles of the creation and annihilation operators of one oscillator. The fit of the mode frequencies yields coupling rates of $g/2\pi= 775 \pm 33~\text{Hz}$. Similarly, we observe an avoided crossing at $\Delta\omega\approx 2\bar{\Omega}$ as a signature of beamsplitter dynamics, with a coupling rate of $g/2\pi= 931 \pm 52~\text{Hz}$ (Fig. \ref{fig2:scans}d). We consistently obtain lower coupling rates for two-mode squeezing as thermal fluctuations blur the transition into the parity-time symmetry-broken regime \cite{Reisenbauer2024}. 

To further characterize the beamsplitter interaction, we measure the coherent energy exchange between the particles over time in both reciprocal and anti-reciprocal configurations. We trap one neutral particle and one charged particle, allowing us to control the charged particle's variance via electric feedback while suppressing their electrostatic interaction. We apply feedback cooling to the charged particle for $25$ ms to generate an energy difference. At time $t=0$, we switch off the feedback cooling and simultaneously turn on the optical interaction for $25$ ms by adjusting the optical detuning. Consequently, the particles begin to coherently exchange energy (Fig. \ref{fig3:extremes}a), analogous to Rabi oscillations in two-level systems. The exchange frequency is determined by the real part of the frequency difference between the normal modes
\begin{align}
    \Delta\Omega_\mathrm{det} &= \sqrt{(\Delta\Omega - \Delta\omega)^2 + g^2e^{-2ikd}}    \label{Rabifreqs1}
    \\
    \Delta\Omega_\mathrm{sum} &= \sqrt{(2\bar{\Omega} - \Delta\omega)^2-g^2e^{2ikd}},
    \label{Rabifreqs2}
\end{align}
as derived from Eqs. \eqref{eq:RWAdynamics} and \eqref{eq:RWAdynamics2} for reciprocal (with $kd=n\pi$ and $\Delta\omega\approx\Delta\Omega$) and anti-reciprocal (with $kd=(2n+1)\pi/2$ and $\Delta\omega\approx 2\bar\Omega$) interactions, respectively. The mode variances decay toward stationary values at a rate dominated by the mechanical damping from the background gas. Even though the anti-reciprocal interaction is of a dissipative nature, the particles' variances oscillating out-of-phase show that there is still a coherent energy exchange within the system, as expected for a beamsplitter-type interaction. This measurement also demonstrates the potential of this platform to apply instantaneous control protocols to collective states.

We turn to characterizing the two-mode squeezing interaction by reconstructing the stationary phase-space dynamics under drive. We demodulate the particle motion at half the optical detuning $\Delta\omega/2$ ($ \pm\Delta\omega/2$) in the case of reciprocal (anti-reciprocal) interaction, which brings the position records into a rotating frame and allows us to extract particle momenta \cite{EkertKnight, SI}. We plot the correlations of the scaled positions $\tilde{z}_{1,2}$ and momenta $\tilde{p}_{1,2}$, calibrated to the thermal states of non-interacting particles (Fig. \ref{fig3:extremes}b). To avoid unstable amplification, we keep the coupling rate $g/2\pi=360\pm 9$ Hz below the damping rate $\gamma/2\pi=473\pm 32$ Hz. As expected, the particle positions and momenta for reciprocal interaction are correlated and anti-correlated, respectively. However, in the anti-reciprocal case, the positions and momenta are simultaneously correlated, which is typically observed when coupling positive- and negative-mass oscillators. We plot the variances of the collective positions $\tilde{z}_\pm=\tilde{z}_1\pm\tilde{z}_2$ as a function of the optical detuning $\Delta\omega$ for anti-reciprocal interactions, thus confirming dissipative thermal-noise squashing (Fig. \ref{fig3:extremes}c). We extract a parametric squeezing gain $s\approx 0.75$ from the theory fit, in excellent agreement with the theoretical value of $s=g/\gamma\approx 0.76$, and a squeezing of $-2.4$ dB below the thermal noise \cite{SI}. Stronger squashing can be achieved by feedback stabilization of the anti-squeezed quadratures \cite{SchliesserSqueezing}.

\begin{figure}[t!]
    \centering
    \includegraphics[width=1\linewidth]{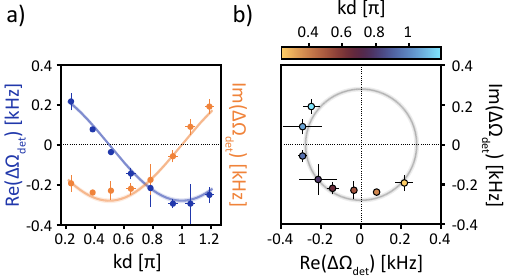}
    \caption{\textbf{Tuning eigenfrequencies in the complex plane.} a) The difference between the real part of the normal mode frequencies (blue) and imaginary part (orange) as a function of the scaled interparticle distance $kd$. The lines are a fit to the theory. The error bars are a combination of errors of the distance, deviations of $\Delta\omega$ from $\Delta\Omega$, and the fit errors. b) The reconstructed phase space of complex eigenfrequencies shows continuous tuning of both the real and imaginary parts that traces a circle around the origin, as predicted by theory (gray line).}
    \label{fig4:arbitrary}
\end{figure}

Up to now, we have focused on purely reciprocal and anti-reciprocal interactions, where changing the distance $kd$ by $\lambda/4$ while keeping $\Delta \omega=\Delta \Omega$ flips their character from beamsplitter to two-mode squeezing. We now scan the intermediate distances and show that the resulting coupling $g_{\text{gen}}(t)$ is a tunable combination of the two operations. This leads to changes in both the mode frequencies and the damping rates, as given by Eq. \eqref{Rabifreqs1}, a scenario that has been inconceivable within the existing toolbox of stationary interactions. We fit the spectra of reconstructed normal modes within the rotating frame with Breit-Wigner functions to extract the differences in frequencies $\mathrm{Re}(\Delta\Omega_\mathrm{det})$ and damping rates $-2\mathrm{Im}(\Delta\Omega_\mathrm{det})$ as a function of the distance (Fig. \ref{fig4:arbitrary}), showing an excellent correspondence to the theory. We note that there are no exceptional points present in Fig. \ref{fig4:arbitrary}b), as we have kept the detuning constant. However, exceptional points exist for other $\Delta \omega$, and they can be encircled by simultaneously varying the detuning and the distance. Therefore, time-dependent interactions can be used to study non-Hermitian effects arising from the continuous tuning of eigenfrequencies, such as mode braiding \cite{Patil2022}.

In summary, we have demonstrated full control over the beamsplitter, single- and two-mode squeezing operations in a pair of optically trapped dipoles -- silica nanoparticles -- with different mechanical frequencies, thereby establishing a complete quantum-optics toolbox for time-dependent light-induced dipolar interactions. This has significant potential for the mechanical control of many-particle systems: (i) the mechanical frequency detuning allows for easier readout and reconstruction of collective modes by filtering in the frequency domain; (ii) beamsplitter, single-mode, and two-mode squeezing operations are distinguishable by controlling the optical detuning, enabling instantaneous protocols to realize collective quantum states; (iii) the mechanical frequency shift that occurs in stationary interactions disappears due to modulation, thus the frequencies remain the same regardless of interaction type or strength \cite{SI}; and (iv) we can realize a negative-mass-like oscillator through anti-reciprocal interactions. We emphasize that points (i) and (ii) are useful even for electrostatic \cite{PoddubnyNonequilibrium} and other modulated interactions. Furthermore, we make the reciprocal beamsplitter and anti-reciprocal two-mode squeezing effects simultaneously resonant, tuning the ratio of their magnitudes via the interparticle distance. This provides a way to jointly tune the normal mode frequencies and damping rates, thereby enabling encirclement of an exceptional point \cite{XuHarris2016, Patil2022}. 

In the anti-reciprocal case, all the signatures of conservative interactions with reversed Stokes and anti-Stokes sidebands for a single degree of freedom suggest an analogy to negative-mass oscillator dynamics, at least in the classical domain. Negative mass oscillators in the quantum regime enable continuous measurement of conjugate system quadratures simultaneously beyond the Heisenberg limit, thereby entangling the original oscillators without feeding the measurement-induced noise back into the subsystem \cite{tsang_evading_2012, PolzikEntanglement, Silanpää_2021}. The programmable nonreciprocal interactions studied in this work can be applied as reservoir engineering to create stationary entanglement \cite{clerk_lecture_notes_2022, metelmann_non_reci_entanglement_2023, liu_engineering_2021} and investigate topological phases \cite{Fang2017, SlimKitaev}. Moreover, it establishes a platform to investigate the effects of nonreciprocal interactions in heat transfer \cite{HarrisCooling, loos_nonreciprocal_2023}, phase transitions \cite{Fruchart2021, Hanai}, many-body physics \cite{FruchartVitelliReview}, and synchronization \cite{BrunelliNonreciprocalSync}. Finally, Floquet engineering of light-induced interactions could open new research avenues for studying out-of-equilibrium dynamics \cite{FloquetNatureReview, BukovFloquetReview, EckardtReview, GoldmanFloquet}, now in nonreciprocal systems.

\textbf{\textit{Acknowledgments.}} We thank Ewold Verhagen and Philipp Treutlein for insightful discussions. This research was funded in whole or in part by the Austrian Science Fund (FWF) projects 10.55776/STA175 and 10.55776/PAT8785024. For the purpose of Open Access, the author has applied a CC BY public copyright license to any Author Accepted Manuscript (AAM) version arising from this submission. This project has received funding from the European Research Council (ERC) under the European Union’s Horizon 2020 research and innovation programme (grant agreement No 101219400). M.A. acknowledges support from the European Union's Horizon 2022 research and innovation programme under the Marie Sklodowska-Curie grant NEOVITA (grant agreement ID: 101109773). B.A.S. acknowledges funding by the Carl Zeiss Foundation through the project QPhoton and by the DFG-510794108

\textbf{\textit{Author contributions.}} L.E., M.A., M.R., and I.C. designed and built the experiment. L.E. and M.A. performed the measurements and analyzed the data. L.E., M.A., B.A.S., and U.D. developed the theoretical model. U.D. conceived the experiment and supervised the experimental work. All authors were involved in writing and editing the paper.

\bibliographystyle{apsrev4-1}

%

\makeatother

\setcounter{figure}{0}
\setcounter{equation}{0}
\renewcommand{\thefigure}{S\arabic{figure}}
\renewcommand{\theequation}{S\arabic{equation}} 

\clearpage
\appendix

\section{\large Supplementary Materials}

\subsection{Trap generation}

\begin{figure}[ht]
    \centering
    \includegraphics[width=\linewidth]{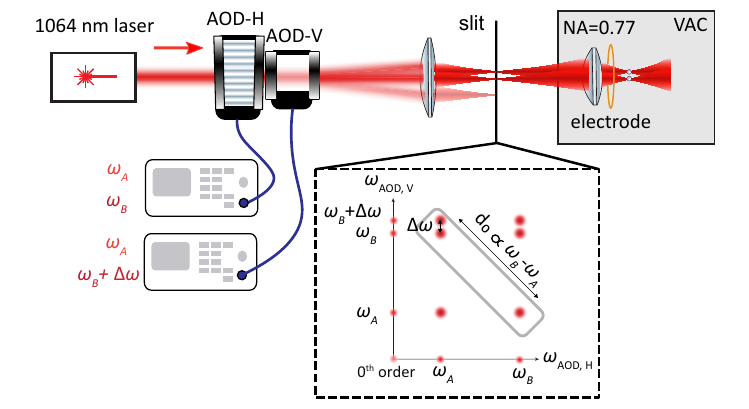}
    \caption{\textbf{The experimental setup to generate detuned optical tweezers.} A $1064$ nm laser beam gets diffracted on a horizontal (AOD-H) and vertical (AOD-V) AOD successively. A slit blocks the off-diagonal traps, the diagonal ones are reimaged onto a trapping lens inside a vacuum chamber. The electrode next to the trapping lens is used for feedback cooling. Tuning $\Delta\omega$ lets us change the modulation frequency through the tweezer detuning while tuning $\Delta\omega_{\textrm{AOD}}= \omega_B-\omega_A$ changes the interparticle distance while keeping the modulation frequency constant.}
    \label{figS1:setup}
\end{figure}

The optical tweezers are created using an acousto-optic deflector (AOD), driven by different radio-frequency (RF) tones to obtain non-degenerate frequencies. We first drive an acousto-optic deflector (AOD) aligned with the horizontal axis of the lab frame with two radio-frequency (RF) tones $\omega_{H,A}$ and $\omega_{H,B}$, creating two diffracted laser beams from the incident $1064$ nm continuous-wave (CW) laser source. They are reimaged via a 4f telescope into a second, vertically aligned AOD, driven with tones $\omega_{V,A}, \omega_{V,B}$, thus creating a $2\times 2$ trap array, as shown in Figure \ref{figS1:setup}.

If the drive tones of the two AODs are chosen to be related as $\omega_{H,B}=\omega_{V,B}$ and $\omega_{H,A}=\omega_{V,A}$, then the optical frequencies of the diagonal traps in the array are resonant. This configuration has been used to explore the position-position coupling between particles of approximately degenerate frequencies \cite{Reisenbauer2024}. In this work, we introduce a small detuning between, e.g., $\omega_{H,B}$ and $\omega_{V,B}$, such that $\omega_{H,A} = \omega_{V,A} = \omega_{A}$ and $\omega_{V,B} = \omega_{H,B} + \Delta\omega$, while keeping $\omega_{H,B}$ constant. This results in a trap optical detuning of $\Delta\omega$ and can be precisely tuned through an arbitrary waveform generator (AWG, Spectrum Instrumentation M4i.6631-x8). 

The driving powers of the RF tones determine the AOD diffraction efficiency, thereby allowing us to control the trap stiffness and the particle's mechanical frequencies. To change the distance $d$ between the traps, we vary the RF frequency difference $\Delta\omega_{\textrm{AOD}}= \omega_{B}-\omega_{A}$ of the tones driving each AOD symmetrically around the mean frequency. Consequently, the diffraction angle changes, but the overall trap detuning stays the same. After the AODs, a lens translates the angle difference between the different first diffraction orders into a distance between the laser beams in the conjugate plane. We select the desired near-resonant laser beams by placing a slit that blocks the remaining beams. Finally, we re-image this plane, rotated by $45^\circ$ with a Dove prism, into the trapping plane. A high numerical aperture trapping lens  ($\mathrm{NA}=0.77$) is mounted inside a vacuum chamber at a controllable pressure between $10^{-1}$ and $1$ mbar and at room temperature. Under these conditions, the dominant noise contribution to the particles' motion comes from the surrounding gas. The particles are dispersed from a nebulizer (Omron U100) into the vacuum chamber until two particles of roughly equal radii are trapped.

\subsection{Table of experimental parameters}

\begin{table}[h!]
    \centering
    \begin{tabular}{|l|c|c|c|c|}
        \hline
        \makecell{Figure in the\\ main text} & $\gamma/2\pi$ [Hz] & $g/2\pi$ [Hz] & $\frac{|\delta|}{2\pi}$ [Hz]* & $kd$ [$\pi$]\\
        \hline
        3a (top) & $106\pm4$ & $724\pm4$ & $202\pm6$ & \makecell{$-0.029$\\$\pm 0.002$} \\
        \hline
        3a (bottom) & $147 \pm 5$ & $722 \pm 7$ & $318\pm9$ & \makecell{$0.489$\\$\pm0.002$} \\
        \hline
        3b (top) & $570\pm40$ & $806\pm24$ & $670\pm180$ & \makecell{$-0.07$\\$\pm 0.06$} \\
        \hline
        3b (bottom)** & $474\pm 10$ & $357 \pm 10$ & $45\pm122$ &\makecell{$0.51$\\$\pm 0.02$} \\
        \hline
        4 & $466\pm 32$ & $276 \pm 7$ & *** & ****\\
        \hline
    \end{tabular}
\raggedright\newline
* Here, $\delta$ corresponds to a generalized detuning away from resonance, with $|\delta|=|\Delta\Omega-\Delta\omega|$ everywhere but in Fig. 3a (bottom), where $|\delta|=|2\bar\Omega-\Delta\omega|$ \newline\indent
** Fig. 3b (bottom) corresponds to one of the datapoints within Fig. 3c ($\Delta\omega = 4.4$ kHz), hence the same experimental parameters hold for Fig. 3c.
\newline\indent
*** $127\pm87, 155\pm 184, 63\pm98, 198\pm139, 48\pm158, 74\pm138, 2\pm122, 142\pm92$
\newline\indent
**** $1.19\pm 0.03, 1.06\pm 0.04, 0.94 \pm 0.03,  0.78\pm 0.02, 0.65\pm 0.04, 0.51\pm 0.02, 0.39 \pm 0.01, 0.24\pm 0.02$ \newline\indent
The values and errors obtained for $|\delta|/2\pi$ come from the mean and standard deviation of $\Omega_1$ and $\Omega_2$ averaged over a range of many kHz of optical detuning $\Delta\omega$. The remaining values and errors correspond to fitting results and errors.
\label{tab:experimental_params}
\end{table}

\subsection{Time-dependent optical binding}\label{sec:S-theory}

In this section, we derive the equations of motion for different resonant cases of the time-dependent interaction between two optically trapped nanoparticles. The optical binding that couples two (identical) nanoparticles at positions $\mathbf{r}_j$, $j=1,2$, stems from the interference between the tweezer electric fields and the fields scattered off both particles. The resulting axial (along the $z$-axis) force acting from particle $j'$ onto particle $j\neq j'$ can be derived from the modification of the optical potential for the $j$-th particle \cite{Rieser2022}
\begin{equation}
    F^\mathrm{bind}_{j,z}=\frac{k^2\alpha^2\sin^2(\theta)}{2\varepsilon_0}  \Re\left[
    \frac{\partial}{\partial z} \left(E^*_j(\mathbf{r}_j) E_{j'}(\mathbf{r}_{j'}) \frac{e^{ikR}}{4\pi R}
     \right) \right],
\end{equation}
where $R=|\mathbf{r}_1-\mathbf{r}_2|$ is the interparticle distance, $\alpha=3\varepsilon_0 V (\varepsilon-1)/(\varepsilon+2)$ is the particle polarizability with volume $V$ and relative permittivity $\varepsilon$, $\theta$ is the laser polarization angle with respect to the direction connecting the particles, and $k$ is the laser wavenumber. As in the previous studies, the tweezer electric field $E_{j}=|E_{0,j}|e^{-i(\omega_{j}t+\phi_{j})}$ can be tuned through magnitudes $|E_{0,j}|$ and phases $\phi_j$. In contrast to previous studies, the tweezer electric fields here are also explicitly functions of the optical frequencies $\omega_j$. Considering only the axial motion along the $z$ axis and denoting the displacement from the trapping minima by $z_j$, respectively, we find in leading order
\begin{align}
&F^{\mathrm{bind}}_{j,z}(t)=  \nonumber\\
&\frac{k^2k'\alpha^2\sin^2(\theta)}{8\pi \varepsilon_0 d}  \left| E^*_{0,j}E_{0,j'} \right|\sin\left[\varphi_j(t) + k'(z_{j'}-z_j)\right], 
\end{align}
where $d$ is the distance between the trap minima, $k'=k-1/z_\mathrm{R}$ is the wavenumber modified by the Rayleigh length $z_\mathrm{R}$, and $\varphi_{1,2}(t)=kd\mp(\Delta\phi +\Delta\omega t)$  is the interference phase for particles placed at the equilibrium positions. This phase is a combination of the traveling phase $kd$, the optical phase difference $ \Delta\phi = \phi_2-\phi_1$ and their accumulated phase due to the optical detuning $\Delta\omega=\omega_2-\omega_1$.

\begin{figure}[t!]
    \includegraphics[width=0.8\linewidth]{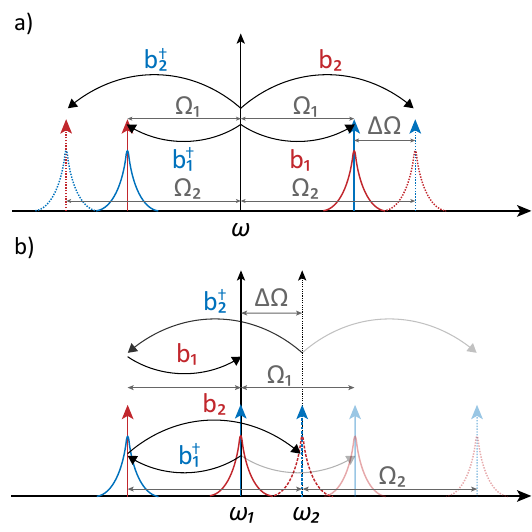}
    \caption{\textbf{Sideband picture of the reciprocal interaction between the particles and light for non-degenerate particle frequencies $\Omega_{1,2}$.} The red (blue) arrows indicate the Stokes (anti-Stokes) sideband created through the creation (annihilation) of a phonon of particle 1 (solid) and particle 2 (dashed). a) For degenerate optical frequencies, the created sidebands are separated by $\Delta\Omega$, making the coupling off-resonant. b) In the case of nondegenerate optical frequencies and $\Delta\omega=\Delta\Omega$, the Stokes sidebands of particle 1 and particle 2 overlap. Therefore, the process where the Stokes-scattered sideband of particle 1 (creation of a phonon) becomes resonant with $\omega_2$ if a phonon of particle 2 is annihilated. The inverse process is also resonant. In total, this yields a beamsplitter interaction.}
    \label{figSM:sideband_picture}
\end{figure}

We linearize the optical binding force with respect to the motion $z_j$. Denoting the local mechanical frequencies by $\Omega_{1,2}$, we find
\begin{equation}
\frac{F^{\mathrm{bind}}_{j,z}}{m\sqrt{\Omega_1\Omega_2}} = 2 g  \left[k^{-1}\sin \varphi_j +
    \cos\varphi_j(z_{j'}-z_{j})\right],\label{eq:Stheoryforce}
\end{equation}
where $m$ is the nanoparticle mass and $g$ is the maximum magnitude of the coupling rate
\begin{equation}
    g=\frac{k^2k'^2\alpha^2\sin^2(\theta)}{16\pi \varepsilon_0 md}\frac{|E^*_{0,j}E_{0,j'} |}{\sqrt{\Omega_1\Omega_2}}.
\end{equation}
The first term in Eq. \eqref{eq:Stheoryforce} is independent of the particle motion and oscillates in time, thus contributing a time-dependent force to the total radiation pressure force. It becomes nonzero on average only when the optical detuning is zero. The second term couples the particle motion and can be tuned via $\varphi_j(t)$. The linearized equation of motion for the $j$-th particle position $z_j$ can be written as
\begin{equation}
    m\ddot z_j+m\gamma \dot z_j+m\Omega_j^2z_j=
    F^{\mathrm{bind}}_{j,z}(t) +F^{\mathrm{rp}}_j+F^\mathrm{th}_j
    \label{eq:EOM_pos}
\end{equation}
where $\gamma$ is the damping rate, which is equal for both particles, $F^\mathrm{rp}_j$ is the radiation pressure force, and $F^\mathrm{th}_j$ is the thermal noise force with $\langle F^\mathrm{th}_j(t)F^\mathrm{th}_{j'}(t')\rangle=2\gamma m k_B T \delta(t-t')\delta_{jj'}$.

\begin{figure*}[ht!]
    \centering
    \includegraphics[width=\textwidth]{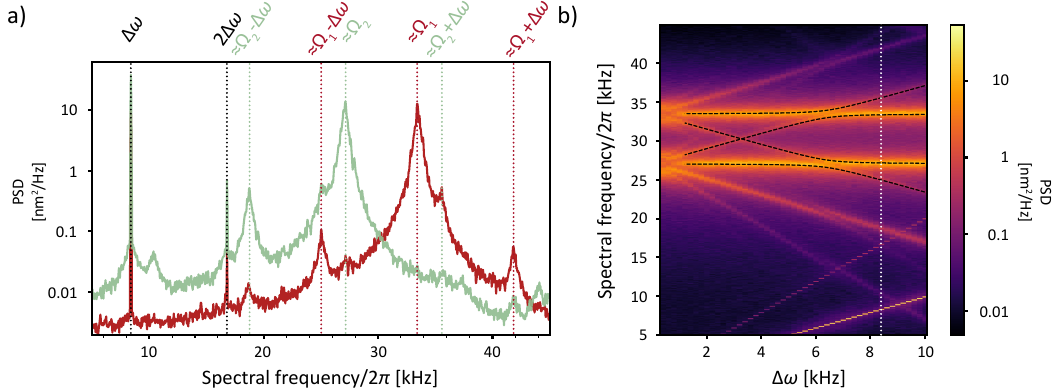}
    \caption{\textbf{Spectroscopy of time-dependent interactions.} 
    a) Experimental power spectral densities (PSDs) of the individual particles corresponding to $\Delta\omega/2\pi=8.4$ kHz in case of reciprocal interaction.
    The modulation of the interaction splits the interaction term in Eq. \eqref{eq:EOM_ai} into sidebands symmetrically shifted by $8.4$ kHz around the particle frequencies $\Omega_1$ and $\Omega_2$.
    The narrow peak at $\Delta\omega/2\pi=8.4$ kHz corresponds to the position-independent part of the optical binding force in Eq. \eqref{eq:Stheoryforce}.
    b) Experimental spectrogram (calculated as the sum of the individual PSDs for an optical detuning scan between $0.2$ kHz and $10$ kHz), showing the dependence of the interaction sidebands on the optical detuning, leading to a resonant interaction between one of them when the optical detuning is close to the mechanical detuning ($\Delta\omega\approx\Delta\Omega$). The white dashed line corresponds to the vertical cut shown in a).}
    \label{figSM:theory}
\end{figure*}

For $\Delta\omega=0$, the equations of motion in Eq. \eqref{eq:EOM_pos} recover the normal mode splitting for mostly reciprocal \cite{Rieser2022} or normal mode attraction for mostly anti-reciprocal interaction \cite{Liska2024, Reisenbauer2024}. Through interaction with light, the particle motion induces Stokes and anti-Stokes sidebands for each tweezer field, which do not interfere for a large difference in the mechanical frequencies $\Delta\Omega$. However, a nonzero optical detuning creates frequency-shifted mechanical sidebands for each particle, enabling the tuning of any pair of sidebands into resonance (Fig. \ref{figSM:sideband_picture}).

To show this in our model, we break down the particle positions into mechanical sidebands $a_je^{-i\Omega_j t}$ and $ a^*_je^{i\Omega_j t}$
\begin{equation}
    z_j=\sqrt{\frac{\hbar}{2m\Omega_j}} \left(a_je^{-i\Omega_j t}+a^*_je^{i\Omega_j t}\right)
    \label{eq:def_compl_amp}
\end{equation}
with slowly varying complex amplitudes $a_j$. We substitute Eq. \eqref{eq:def_compl_amp} into Eq. \eqref{eq:EOM_pos} and use the slowly varying envelope approximation under the assumption of high quality factor oscillators ($|\ddot a_j|\ll|\Omega_j \dot a_j|$ and $|\dot a_j|\ll|\Omega_j a_j|$) to get
\begin{widetext}
\begin{equation}
     \dot a_j+\frac{\gamma}{2} a_j  =ig\cos \varphi_j
    \left(\left(a_{j'}e^{i(-1)^j\Delta\Omega t}+a^\dagger_{j'}e^{2i\bar\Omega t}\right) -\sqrt{\frac{\Omega_{j'}}{\Omega_j}} \left(a_j+a^\dagger_j e^{2i\Omega_j t}\right)\right)+a_{j,\mathrm{in}},
\label{eq:EOM_ai}
\end{equation}
\end{widetext}
where we defined the mechanical detuning $\Delta\Omega = \Omega_2-\Omega_1$, the mean mechanical frequency $\bar\Omega=(\Omega_1+\Omega_2)/2$, and the input noise amplitude $a_{j,\mathrm{in}}=iF^\mathrm{th}_j e^{i\Omega_j t} \sqrt{1/(2m\hbar\Omega_j)}$ in units of angular frequency. By expressing $\cos\varphi_j(t)$ in the complex form
\begin{align}
&\cos \left[kd\pm(\Delta\phi +\Delta\omega t)\right]=\nonumber
\\
&\frac{1}{2}\left(e^{i(kd\pm \Delta\phi)}e^{\pm i\Delta\omega t}+e^{-i(kd\pm \Delta\phi)}e^{\mp i\Delta\omega t}\right), 
\end{align}
we see that terms in Eq. \eqref{eq:EOM_ai} become effectively time-independent if $\Delta\omega$ is resonant with $\pm\Delta\Omega$, $\pm 2\bar{\Omega}$, or $\pm 2\Omega_j$. For $\Delta\omega=0$, the term proportional to $\cos \varphi_j a_j$ modifies the mechanical frequency through optical binding. However, for $\Delta\omega\neq 0$, this term averages to zero and will be neglected in the following text. In Figure \ref{figSM:theory}, we show experimental power spectral densities of the particle motion and mark all relevant sidebands that stem from the time-dependent interactions.

\subsection{Eigenmode reconstruction}

Here, we assume nondegenerate mechanical frequencies ($\Delta\Omega\neq 0$) and a nonzero optical detuning ($\Delta\omega\neq 0$), and solve the system dynamics at the different interaction resonances.

\paragraph{Optical detuning close to mechanical detuning.}
In the case of $\Delta\omega\approx \Delta\Omega$, the term proportional to $a_{j'}$ dominates the system dynamics, while the other terms can be neglected within the rotating wave approximation (RWA), as the optical detuning is far from $2\bar{\Omega}$ or $2{\Omega}_{1,2}$. The dynamics of $a_j$ thus simplify to
\begin{equation}
\dot a_j+\frac{\gamma}{2} a_j  =
i\frac{g}{2} e^{-i\left(kd+(-1)^j\Delta\phi\right)}e^{-i(-1)^j(\Delta\omega-\Delta\Omega) t}  a_{j'}+
  a_{j,\mathrm{in}}.
\label{eq:EOM_dfd0}
\end{equation}
In comparison to the case of $\Delta\omega=0$, the coupling is now reduced by a factor of two; this is a consequence of the RWA discarding the off-resonant sideband of the sinusoidal drive at $\Delta\omega+\Delta\Omega\approx 2\Delta\Omega$. We note that the other sideband can be chosen by changing the sign of the optical detuning to $\Delta\omega=-\Delta\Omega$. To eliminate the time-dependence of the coupling rate in Eq. \eqref{eq:EOM_dfd0}, we perform a transformation of the complex amplitudes $b_{1,2}= a_{1,2} e^{\mp i(\Delta\omega-\Delta\Omega) t/2}$. Effectively, this moves the complex amplitudes into the rotating frame at the interaction frequency. The equations of motion for $b_j$ are
\begin{align}
    &\left(\frac{\partial}{\partial t}+\frac{\gamma}{2}\right)
    \left(
    \begin{matrix}
        b_1 \\
        b_2
    \end{matrix}
    \right)=
    \nonumber\\
    &\frac{i}{2}
    \left(
    \begin{matrix}
        \Delta\Omega-\Delta\omega
        &
        \tilde g\textrm{e}^{-ikd} 
        \\
        \tilde g ^*\textrm{e}^{-ikd }
        &
        \Delta\omega-\Delta\Omega
    \end{matrix}
    \right)
    \left(
    \begin{matrix}
        b_1 \\
        b_2
    \end{matrix}
    \right)+
    \left(
    \begin{matrix}
        b_{1,\mathrm{in}} \\
        b_{2,\mathrm{in}}
    \end{matrix}
    \right),    
    \label{eq:EOM_dfd}
\end{align}
where we substituted $\tilde{g} = g e^{i\Delta\phi}$.

\begin{figure}[t]
    \includegraphics[width=\linewidth]{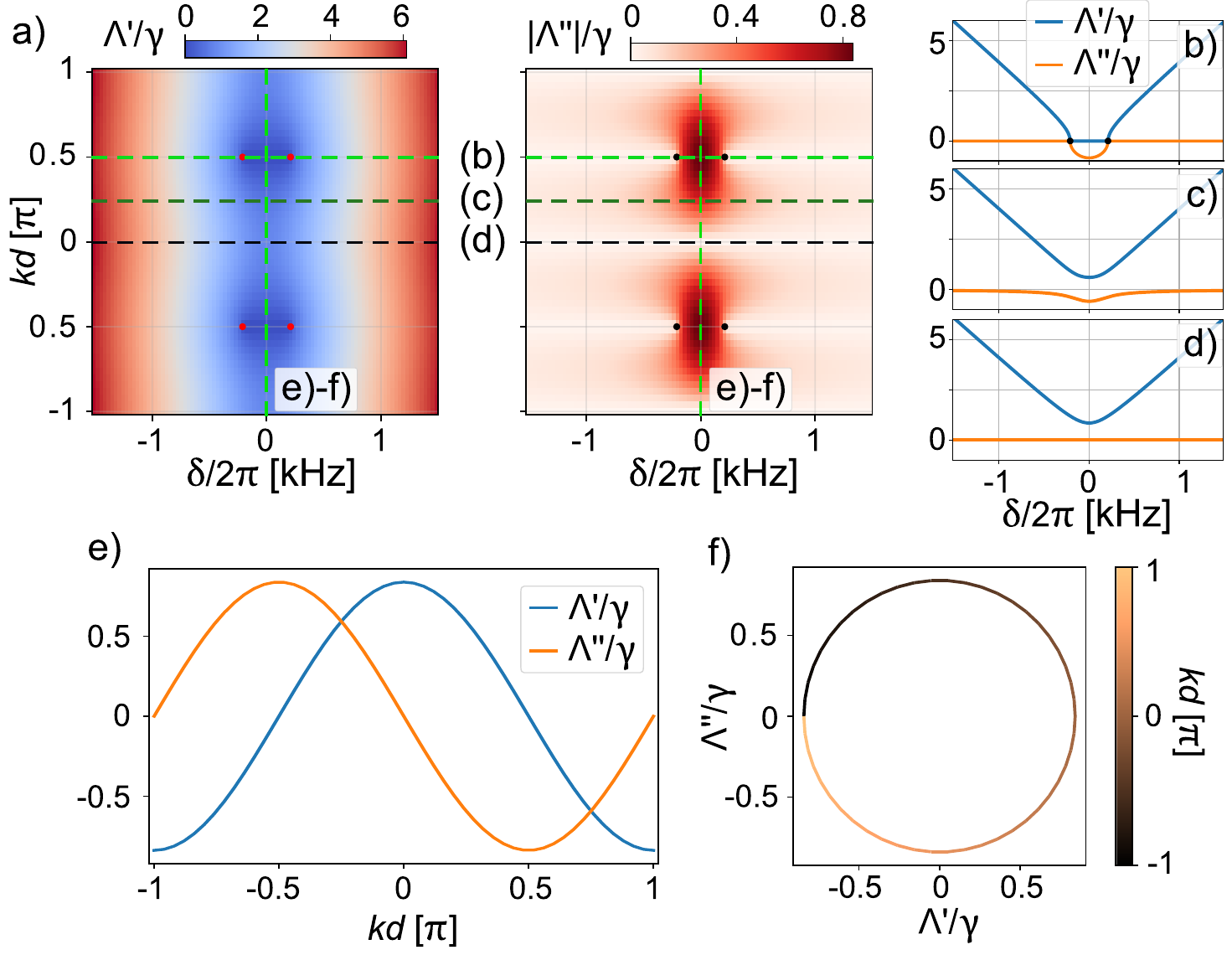}
    \caption{\textbf{Eigenfrequencies around the resonant interactions.}
    a) Color maps of the real ($\Lambda'=\Re(\Lambda_{\mathrm{det}})$) and imaginary ($\Lambda''=\Im(\Lambda_\mathrm{det})$) parts of the eigenfrequency as a function of the effective detuning and distance.
    The red and black points at $kd=(2n+1)\pi/2$ are the exceptional points.
    b)-d) Horizontal cuts from a) showing both the real and imaginary parts of the eigenfrequency as a function of the effective detuning at $kd=(2n+1)\pi/2$, $kd=n\pi+\pi/4$, and $kd=n\pi$, respectively.
    e) Vertical cuts from a) showing both the real and imaginary parts of the eigenfrequency as a function of the distance $kd$ at zero effective detuning.
    f) Trajectory of the resonant ($\delta=0$) eigenfrequency in the $\Lambda'=\Re(\Lambda_\mathrm{det})$ and $\Lambda''=\Im(\Lambda_\mathrm{det})$ plane as function of the distance.
}
    \label{figSM:eigs}
\end{figure}

The eigenfrequencies and right eigenvectors of Eq. \eqref{eq:EOM_dfd} are
\begin{align}
    \Omega_{\mathrm{det},\pm}&=\pm \frac{1}{2} \sqrt{ (\Delta\omega-\Delta\Omega)^2 + g^2e^{-2ikd}}:=\pm \frac{1}{2}\Lambda_\mathrm{det}\nonumber
 \\ 
    \mathbf{n}_{\mathrm{det},\pm}&=
    \frac{e^{i\Delta\phi}}{\sqrt{ \mathcal{N}} }
    \left(
    \begin{matrix}
    e^{ikd} \left(-(\Delta\omega-\Delta\Omega)+2\Omega_{\mathrm{det},\pm}\right) \\
    \tilde g^*
    \end{matrix}
    \right)\nonumber
\\
\mathcal{N} &= |-(\Delta\Omega-\Delta\omega) +2\Omega_{\mathrm{det},\pm} |^2+g^2,
\label{eq:dif_det_normalmodes}
\end{align}
where the normalization is recovered from $\mathbf{n}_{\mathrm{det},\pm}^*\cdot \mathbf{n}_{\mathrm{det},\pm}=1$.
The four dashed lines in Fig. \ref{fig2:scans}a) and c) are fit functions $\bar{\Omega}\pm\Delta\omega/2+\Omega_{\mathrm{det},\pm}$ as a function of $\Delta\omega$, from which we obtain $g$.
More generally, the variation of the eigenfrequency, $\Lambda_{\mathrm{det}}$, as a function of the effective detuning $\delta=\Delta\omega-\Delta\Omega$ and the distance phase $kd$ is shown in Fig.\ref{figSM:eigs} for both its real and imaginary part, accounting for the system's conservative and dissipative dynamics.
The eigenfrequency is purely real only for $kd=n\pi$ with $n\in\mathbb{N}$.
For $kd=(2n+1)\pi/2$, the eigenfrequency switches from real to imaginary (and vice versa) at $\delta=-g$ ($\delta=g$), defining the locations of the exceptional points (EP-s) highlighted with red and black points in Fig.\ref{figSM:eigs}. Everywhere else, the eigenfrequency is complex, with non-zero real and imaginary parts.

In Figure \ref{fig4:arbitrary} in the main text, we show how the real and imaginary parts of the frequency difference $\Delta\Omega_\mathrm{det}$ change as a function of $kd$. To measure $\Re(\Delta\Omega_\mathrm{det})$ and $\Im(\Delta\Omega_\mathrm{det})$, we reconstruct the eigenmodes in the frame rotating with $\pm\Delta\omega/2$, which is done using a Hilbert transform $\mathcal{H}$ such that $z_j^\text{RWA} =\Re(\mathcal{H}(z_j)e^{\mp i \Delta\omega t/2})$ (see Methods in \cite{Reisenbauer2024}). Using  $(-1)^j\Delta\Omega/2=\Omega_j-\bar\Omega$, we remark that the operators $b_j e^{-i\bar\Omega t}$ are still in a frame with constant coupling because the global frequency shift does not affect the collective dynamics. 
These are the observables that we calculate experimentally from the motional sidebands (Hilbert transforms of position) by demodulating them at half the optical detuning, with the sign given by $(-1)^j$
\begin{equation}
    b_j e^{-i\bar\Omega t}=(a_j e^{-i\Omega_j t}) e^{i(-1)^j \Delta\omega t/2 }.
\end{equation}

\begin{figure}[t!]
    \includegraphics[width=\linewidth]{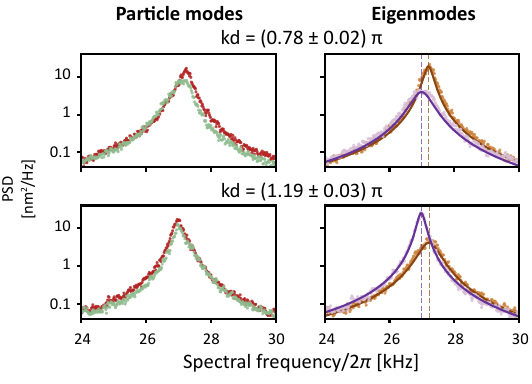}
    \caption{\textbf{Power spectral densities (PSD) of the particle modes and their reconstructed eigenmodes for different interparticle distances.} The left plot shows the PSD of the individual particles (green and red) for two different interparticle distances for which both reciprocal and anti-reciprocal couplings are present. The right column shows the PSD of the particles' reconstructed eigenmodes $\mathbf{n}_\pm$ (purple and brown) at the same interparticle distances, along with their fit. The purple mode corresponds to the sum of the demodulated and phase corrected particle motion, and the brown mode to their difference. The points are data, and the lines are fits to a Breit-Wigner function. The dashed vertical lines mark the extracted real parts of the eigenfrequencies. As $kd$ changes from $kd<\pi$ to $kd>\pi$, the role of the over- and underdamped mode flips, since $\Delta\gamma$ is proportional to $\sin(kd)$. However, the individual particle PSDs remain roughly the same.} 
    \label{figS:nm-reconstruction}
\end{figure}

We set the optical detuning such that $\Delta\omega\approx \Delta\Omega$, so we can approximate the eigenvectors and eigenfrequencies from Eq. \eqref{eq:dif_det_normalmodes} as
\begin{equation}
    \mathbf{n}_\pm=\frac{1}{\sqrt{2}}
    \left(
    \begin{matrix}
    \pm e^{i\Delta\phi} \\
    1
    \end{matrix}
    \right), \quad  \Omega_{\mathrm{det},\pm}=\pm \frac{1}{2} g e^{-ikd}.
    \label{eq:normalmodes_rwa}
\end{equation}
The phase difference $\Delta\phi$ corresponds to the sum of the beatnote phase at the start of the measurement $\Delta\omega t_0$ and the tweezer phase difference in the absence of optical detuning, which we compensate for by calculating the phase offset when moving into the rotating frame. For $\Delta\phi=0$, the two eigenvectors correspond exactly to the sum and difference of the particles' motion in this frame. From Eq. \eqref{eq:normalmodes_rwa}, we see that the real and imaginary parts of the difference between the eigenfrequencies move along a closed circle in the complex plane for $kd:-\pi\rightarrow \pi$. However, correctly mapping the eigenfrequencies onto the eigenvectors requires access to the initial phase $\Delta\phi$. 

Since we did not record the initial phase $\Delta\phi$ in this experiment, we instead compute the mean phase difference between the particles' motion in the rotating frame and remove it, such that the contribution of the mode with decreased linewidth is always projected onto the $\mathbf{n}_+$ mode, since its contribution to the particle's phase difference is the largest. As can be seen in Eq. \eqref{eq:normalmodes_rwa}, this is equivalent to setting $\Delta\phi=0$ for $kd\in(-\pi,0)$, where $\Omega_{\mathrm{det},+}$ has a reduced linewidth (since $\gamma = -2\Im(\Omega_\mathrm{det})$); however, for $kd\in(0, \pi)$, $\Omega_{\mathrm{det},-}$ becomes the eigenmode with reduced linewidth. Therefore, this method means that we need to manually flip the definitions of $\mathbf{n}_\pm$ within this region of $kd$ to reconstruct the correct mapping between $\mathbf{n}_\pm$ and $\Omega_{\mathrm{det},\pm}$. After the successful eigenmode reconstruction, we fit the resulting power spectral densities with a Breit-Wigner function to extract the difference of the oscillation frequencies $\Re\left(\Omega_{\mathrm{det},+}\right)-\Re\left(\Omega_{\mathrm{det},-}\right)=\Re(\Delta\Omega_\mathrm{det})$ and damping rates $\Im\left(\Omega_{\mathrm{det},+}\right) - \Im\left(\Omega_{\mathrm{det},-}\right) = \Im(\Delta\Omega_\mathrm{det}) = -(\gamma_{\mathrm{det},+}-\gamma_{\mathrm{det},-})/2$.

\paragraph{Optical detuning close to $2\bar{\Omega}$.}
If the optical detuning is now tuned close to the sum of both mechanical frequencies ($\Delta\omega\simeq 2\bar\Omega$), only the term $\propto a^\dagger_{j'}$ in Eq. \eqref{eq:EOM_ai} is resonant. We apply the RWA again to obtain
\begin{align}
\dot a_j+\frac{\gamma}{2} a_j  \simeq
i\frac{g}{2} e^{-i\left((-1)^j kd+\Delta\phi\right)}e^{i(\Delta\omega-2\bar{\Omega}) t} a^\dagger_{j'}+a_{j,\mathrm{in}}.
\label{eq:EOM_sfd}
\end{align}
To move to a rotating frame where the coupling is constant, we now define $b_j=a_j e^{i(\Delta\omega-2\bar\Omega)t/2}$ and obtain the following equation for $b_j$
\begin{align}
    &\left(\frac{\partial}{\partial t}+\frac{\gamma}{2}\right)
    \left(
    \begin{matrix}
        b_1 \\
        b^\dagger_2
    \end{matrix}
    \right)=
    \nonumber\\
    &\frac{i}{2}
    \left(
    \begin{matrix}
        \Delta\omega-2\bar\Omega
        &
        \tilde g^*e^{ikd} 
        \\
        -\tilde g e^{ikd }
        &
        2\bar\Omega-\Delta\omega
    \end{matrix}
    \right)
    \left(
    \begin{matrix}
        b_1 \\
        b_2^\dagger
    \end{matrix}
    \right)+
    \left(
    \begin{matrix}
        b_{1,\mathrm{in}} \\
        b^\dagger_{2,\mathrm{in}}
    \end{matrix}
    \right)    
    \label{eq:EoM_sfd}
\end{align}
To link these results with experimental data processing, we define $(b_1,b^\dagger_2)\textrm{e}^{i\Delta\Omega t/2}$, which are still in a rotating frame with stationary interactions because the global frequency shift $\Delta\Omega/2$ does not affect the collective dynamics. 
We obtain these operators by demodulating the mechanical sidebands at half the optical detuning
\begin{equation}
    (b_1,b^\dagger_2)e^{i\Delta\Omega t/2}=((a_1e^{-i\Omega_1 t})e^{i\Delta\omega t/2},(a^\dagger_2e^{i\Omega_2 t})e^{-i\Delta\omega t/2}).
\end{equation}

The eigenfrequencies and right eigenvectors of equation \eqref{eq:EoM_sfd} are
\begin{align}
    \Omega_{\mathrm{sum},\pm}&=
    \pm \frac{1}{2} \sqrt{ (\Delta\omega-2\bar\Omega)^2 -  g^2 e^{2ikd}}:=
    \pm\frac{\Lambda_\mathrm{sum}}{2}   
    \nonumber 
    \\
    \mathbf{n}_\pm&=
    \frac{e^{-i\Delta\phi}}{\sqrt{\mathcal{N}} }
    \left(
    \begin{matrix}
    -e^{-i kd} (\Delta\omega-2\bar\Omega+2\Omega_{\mathrm{sum},\pm})  \\
    \tilde g &
    \end{matrix}
    \right) \nonumber \\
    \mathcal{N} &= |\Delta\omega-2\bar\Omega+2\Omega_{\mathrm{sum},\pm}|^2 + g^2.
\end{align}
The four dashed lines in Fig. \ref{fig2:scans}b) and d) in the main text are fitted by four possible combinations $\pm\Delta\Omega/2+\Delta\omega+\Omega_{\mathrm{sum},\pm}$.

\subsection{Single-mode squeezing}

Single-mode squeezing (SMS) interaction can be engineered by setting the optical detuning to twice the mechanical frequency of any particle $\Delta\omega = 2\Omega_j$. Intuitively, this can be understood as the light elastically scattered off one particle, parametrically driving the other. Eq. \eqref{eq:EOM_ai} now shows resonant coupling only for the term $\propto a^\dagger_j$
\begin{equation}
\dot a_j+\frac{\gamma}{2} a_j  \simeq
-i\frac{g_\mathrm{s}}{2} e^{-i\left((-1)^j kd+\Delta\phi\right)}e^{i(\Delta\omega-2\Omega_j) t}  a^\dagger_j+a_{j,\mathrm{in}},
\label{eq:EOM_tfd}
\end{equation}
where we substituted $g_s=g\sqrt{\Omega_{j'}/\Omega_j}$. A rotating frame with constant coupling can be found by multiplying Eq. \eqref{eq:EOM_tfd} by  $e^{i(\Delta\omega-2\Omega_j)t/2}$ and defining $b_i=a_je^{i(\Delta\omega-2\Omega_j)t/2}$ to get the following equation for $b_j$
\begin{align}
    &\left(\frac{\partial}{\partial t}+\frac{\gamma}{2}\right)
    \left(
    \begin{matrix}
        b_1 \\
        b^\dagger_1
    \end{matrix}
    \right)=
    \nonumber\\
    &\frac{i}{2}
    \left(
    \begin{matrix}
        \Delta\omega-2\Omega_i
        &
        -\tilde g^*e^{ikd} 
        \\
        \tilde g e^{ikd }
        &
        2\Omega_i-\Delta\omega
    \end{matrix}
    \right)
    \left(
    \begin{matrix}
        b_1 \\
        b_1^\dagger
    \end{matrix}
    \right)+
    \left(
    \begin{matrix}
        b_{1,\mathrm{in}} \\
        b^\dagger_{1,\mathrm{in}}
    \end{matrix}
    \right).
    \label{eq:EoM_tfd}
\end{align}
The operators $b_j$ correspond to the motional sidebands demodulated at half the optical detuning. The eigenfrequencies and right eigenvectors of equation \eqref{eq:EoM_tfd} are
\begin{align}
    \Omega_{\mathrm{SMS},\pm}&=\pm \frac{1}{2} \sqrt{ (\Delta\omega-2\Omega_j)^2 -  g_\mathrm{s}^2e^{2ikd}} 
 \nonumber \\
    \mathbf{n}_{\mathrm{SMS},\pm}&=
    \frac{e^{-i\Delta\phi}}{\sqrt{\mathcal{N}} }
    \left(
    \begin{matrix}
    e^{-i kd} (\Delta\omega-2\Omega_j+2\Omega_{\mathrm{SMS},\pm}) \\
    \tilde g &
    \end{matrix}
    \right) \nonumber \\
    \mathcal{N} &= |\Delta\omega-2\Omega_j+2\Omega_{\mathrm{SMS},\pm}|^2 + g^2
\end{align}

\begin{figure}[t]
    \includegraphics[width=\linewidth]{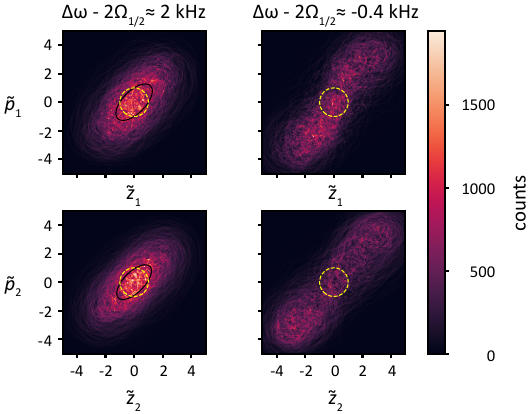}
    \caption{\textbf{Single-mode squeezing through the light-induced dipolar interaction.} Position-momentum phase space of particle 1 (top) and 2 (bottom) scaled by the standard deviation of the thermal state of the unmodified other particle (yellow dashed circles). These measurements correspond to different vertical cuts in Figure \ref{fig2:scans}b. The left plots correspond to a single-mode squeezing interaction within the linear regime. Here, we fit the diagonal and anti-diagonal in order to obtain the black ellipse, which shows a reduced thermal variance with respect to the unperturbed other particle (yellow dashed line). As we move closer to resonance, we observe the motion becoming highly nonlinear (right).} 
    \label{figS:SMS}
\end{figure}

Fig. \ref{figS:SMS} shows the single particle phase-space in the rotating frame within the linear regime (left) and the nonlinear regime (right). In the linear regime, we can see that the single-particle variance is reduced with respect to the variance of the other particle (yellow dashed line) along one of the joint system quadratures, while in the nonlinear regime, we see the emergence of a bistable state, with the particle position displaced by up to $\sim\pm\lambda/2$.

\section{System dynamics}

In all resonant cases, the equations of motion are linear and time-independent, implying that they can be written as
\begin{eqnarray}
    \left( \frac{\partial}{\partial t}+\frac{\gamma}{2} \right)\mathbf{b} = iH\mathbf{b} + \mathbf{b}_\mathrm{in}
    \label{eq:mathbfb}
\end{eqnarray}
where $\mathbf{b}$ is the mode amplitude vector with its components given by the resonantly coupled variables, $H$ is the (time-independent) dynamical matrix describing the coupled evolution, and  $\mathbf{b}_{\mathrm{in}}$ is the thermal noise force vector (in units of angular frequency). 

We solve for $\mathbf{b}$ by explicitly calculating the matrix exponential of $H$. Specifically, its right eigenvectors and eigenvalues allow us to construct the passage matrix
$P = \sum_{i = 1}^3 \mathbf{n}_i \otimes \mathbf{e}_i$ that transforms from the canonical basis $\mathbf{e}_i$ into the right eigenvector basis $\mathbf{n}_i = P \mathbf{e}_i$. Its inverse matrix transforms back into the canonical basis, $P^{-1} \mathbf{n}_j = \mathbf{e}_j$ (implying that it can be given in terms of the dual left eigenvectors). With this, we can write the non-Hermitian dynamical matrix in the form $H = P H_0 P^{-1}$, where $H_0$ is the diagonal matrix of the eigenvalues. It is now straightforward to solve Eq. \eqref{eq:mathbfb} by direct integration, yielding
\begin{equation}
    \textbf{b}(t)=U(t)
    \textbf{b}(0)e^{-\frac{\gamma}{2} t}+
    \int\limits_0^t dt' e^{-\frac{\gamma}{2} (t-t')} U(t - t')\textbf{b}_{\mathrm{in}}(t')
    \label{eq:ttrsol}
\end{equation}
with
\begin{eqnarray}
    U(\tau)=
    P\textrm{e}^{iD \tau}P^{-1}=
    \left(
    \begin{matrix}
    \alpha_1(\tau) & \beta_1(\tau) \\
    \beta_2(\tau) & \alpha_2(\tau)
    \end{matrix}
    \right).
    \label{eq:evo_matr_coef_def}
\end{eqnarray}

Eq \eqref{eq:ttrsol} is the general solution for the linearly coupled mode amplitudes. The first term describes the evolution of the initial state, while the second term represents the response to the thermal noise force, expressed as the convolution of the thermal noise force with the evolution matrix, which serves as the Green's function for the dynamical system under consideration. As $U$ depends only on $t-t'$, this implicitly assumes the time-invariance of the dynamical system, i.e., the coefficients of $H$ are time-independent between $t'=0$ and $t'=t$, which is only true for the modes in a rotating frame.

The coefficients of the evolution matrix are
\begin{align}
    & \alpha_{1,2}(\tau) = 
    \cos\left(\frac{\Lambda_\mathrm{det}}{2}\tau\right)
    \mp i \frac{ \Delta\omega - \Delta\Omega }{\Lambda_\mathrm{det}} 
    \sin\left(\frac{\Lambda_\mathrm{det}}{2}\tau\right)
    \nonumber
    \\
    & \beta_{1,2}(\tau)=\
    i \frac{g e^{\pm i\Delta\phi}}{\Lambda_\mathrm{det}} 
    \sin\left(\frac{\Lambda_\mathrm{det}}{2}\tau\right)
    e^{-ikd }
    \label{eq:detevocoeff4}
\end{align}
for $\Delta\omega\approx\Delta\Omega$ and
\begin{align}
    & \alpha_{1,2}(\tau) = 
    \cos\left(\frac{\Lambda_\mathrm{sum}}{2}\tau\right)
    \pm i\frac{\Delta\omega-2\bar{\Omega}}{\Lambda_\mathrm{sum}} 
    \sin\left(\frac{\Lambda_\mathrm{sum}}{2}\tau\right)
    \nonumber
    \\
    & \beta_{1,2}(\tau)=
    \pm i\frac{g e^{\mp i\Delta\phi}}{\Lambda_\mathrm{sum}} 
    \sin\left(\frac{\Lambda_\mathrm{sum}}{2}\tau\right)
    e^{i kd }
    \label{eq:detevocoeff_sfd}
\end{align}
for $\Delta\omega\approx2\bar{\Omega}$.

\subsection{First-order correlation functions}

In this section, we derive the first-order correlation functions
\begin{eqnarray}
    & g_{jj'}^{(1)}(t,\tau)=\langle b^*_j(t)b_{j'}(t+\tau)\rangle 
    \label{eq:g1_def}
\end{eqnarray}
where $\langle\cdots\rangle$ is the ensemble average. In the stationary case, the correlations only depend on the relative time delay $\tau$ due to time invariance. We apply the solutions from Eq. \eqref{eq:ttrsol} into Eq. \eqref{eq:g1_def}. Taking the limit $t\rightarrow\infty$, we make the correlations lose the memory of the system's initial state and therefore neglect the term depending on the initial conditions.
We use $\langle b^*_{j,\mathrm{in}}(t) b_{j',\mathrm{in}}(t')\rangle=
\delta_{jj'}\gamma n_j \delta(t-t')$ with $n_j=k_BT/(\hbar\Omega_j)$ to obtain
\begin{align}
    & \langle b^*_j(t)b_j(t+\tau)\rangle
    =
    \gamma e^{-\frac{\gamma}{2} \tau} \nonumber \\
    & \int\limits_{0}^\infty
    dt' e^{-\gamma t'} 
    \left[
    n_j\alpha^*_j(t')\alpha_j(t'+\tau)+
    n_{j'}\beta_j^*(t')\beta_j(t'+\tau)
    \right] 
    \label{eq:gencorrexp1}
    \\
    & \langle b^*_1(t)b_2(t+\tau)\rangle
    =
    \gamma e^{-\frac{\gamma}{2} \tau} \nonumber \\
    &\int\limits_{0}^\infty
    dt' e^{-\gamma t'}
    \left[n_1
    \alpha^*_1(t')\beta_2(t'+\tau)+
    n_2
    \beta_1^*(t')\alpha_2(t'+\tau)
    \right].
    \label{eq:gencorrexp3}
\end{align}
In the equations above, the time delay is assumed to be positive ($\tau>0$). 
Eqs. \eqref{eq:gencorrexp1} and \eqref{eq:gencorrexp3} can be solved analytically to obtain $g^{(1)}_{jj'}(\tau)$.

We first write $g^{(1)}_{12}(\tau)$
as:
\begin{align}
    g^{(1)}_{12}(\tau)&=
    e^{-\frac{\gamma}{2}|\tau|}
    A \cos\left(\frac{\Lambda}{2}\tau-B\right),
    \label{eq:g1_12_simplified}
\end{align}
where $\Lambda$ can be either $\Lambda_\mathrm{det}$ or $\Lambda_\mathrm{sum}$, depending on $\Delta\omega$. Coefficients $A$ and $B$ also generally have different values depending on the sign of $\tau$. To evaluate $g^{(1)}_{12}(-\tau)$, we need to either change the lower integration bound to $-\tau$, or calculate $\langle b^*_1(t+\tau)b_2(t) \rangle$, which amounts to swapping $t'$ and $t'+\tau$ in Eq. \eqref{eq:gencorrexp3}. The coefficients are
\begin{widetext}
\begin{align}
&A=\begin{dcases*}
    2\gamma\tilde{g}^*\sqrt{\frac{n_1\left(2i\gamma-\delta+\Lambda\right)e^{-ikd}
+n_2\left(\Lambda+\delta\right)e^{ikd}}{4i\Lambda(\gamma+\Lambda'')(\gamma-i\Lambda')}
\frac{n_1\left[-(2i\gamma-\delta)+\Lambda\right]e^{-ikd}
+n_2\left(\Lambda-\delta\right)e^{ikd}}{4i\Lambda(\gamma-\Lambda'')(\gamma+i\Lambda')}}\text{, }\tau\geq 0\\
2\gamma\tilde{g}^*\sqrt{\frac{n_2\left(2i\gamma-\delta+\Lambda^*\right)e^{ikd}
+n_1\left(\Lambda^*+\delta\right)e^{-ikd}}
{4i\Lambda^*(\gamma-\Lambda'')(\gamma-i\Lambda')}
\frac{n_2\left[-(2i\gamma-\delta)+\Lambda^*\right]e^{ikd}
+n_1\left(\Lambda^*-\delta\right)e^{-ikd}}{
4i\Lambda^*(\gamma+\Lambda'')(\gamma+i\Lambda')}}\text{, }\tau< 0,
\end{dcases*}
\\
&B=
\begin{dcases*}
-i\,\operatorname{arctanh}
\left[
\frac{
\begin{aligned}
&2\gamma^3 n_1
-2i\gamma\Lambda''\Lambda' n_1 +\delta\left(i\gamma^2+\Lambda''\Lambda'\right)
\left(n_1-e^{2ikd}n_2\right) +\gamma\Lambda^2
\left(n_1+e^{2ikd}n_2\right)
\end{aligned}
}{
\begin{aligned}
&\delta\gamma\Lambda
\left(n_1-e^{2ikd}n_2\right) +\Lambda\Lambda''\Lambda'
\left(n_1+e^{2ikd}n_2\right) -i\gamma^2\Lambda\left(n_1-e^{2ikd}n_2\right)
\end{aligned}
}
\right]\text{, }\tau\geq 0\\
i\,\operatorname{arctanh}
\left[
\frac{
\begin{aligned}
&\left(\gamma^2+i\Lambda''\Lambda'\right)
\left(-i\delta n_1
+e^{2ikd}(i\delta+2\gamma)n_2\right) +\gamma(\Lambda^*)^2
\left(n_1+e^{2ikd}n_2\right)
\end{aligned}
}{
\begin{aligned}
&\delta\gamma\Lambda^* n_1 +i\Lambda^* e^{2ikd}\gamma(i\delta+2\gamma)
 n_2 +\left(-i\gamma^2+\Lambda''\Lambda'\right)
\left(n_1+e^{2ikd}n_2\right)\Lambda^*
\end{aligned}
}
\right]\text{, }\tau<0.
\end{dcases*}
\end{align}
\end{widetext}

\subsection{Two-mode squeezing}

In Figure \ref{fig3:extremes}c) in the main text, we showed a measurement of the squashing of thermal noise due to the anti-reciprocal two-mode squeezing interaction. The variance of the squashed and anti-squashed quadratures is
\begin{align}
    \langle z_{\pm}^2\rangle = 
    \frac{ g^{(1)}_{11}(0) + g^{(1)}_{22}(0) }{2}
     \pm \Re\left[ g^{(1)}_{12}(0) \right].
\end{align}
The sum of the motional variances of particles 1 and 2 is
\begin{align}
     &g^{(1)}_{11}(0) + g^{(1)}_{22}(0) = \frac{2\gamma\delta\Lambda'\Lambda''(n_{1}-n_{2})}{2(\gamma^{2}-\Lambda''^{2})(\gamma^{2}+\Lambda'^{2})}\nonumber\\ 
   & +\frac{\gamma^2(n_{1}+n_{2})\!\left(2\gamma^{2}+\Lambda'^{2}-\Lambda''^{2}+\delta^{2}+g^{2}\right)}{2(\gamma^{2}-\Lambda''^{2})(\gamma^{2}+\Lambda'^{2})}
  ,
\end{align}
where $\delta=\Delta\omega-\Delta\Omega$, $\Lambda = \Lambda_\mathrm{det}$, $\Lambda'=\Re(\Lambda_\mathrm{det}$), and $\Lambda''=\Im(\Lambda_\mathrm{det})$. We extract the maximum achievable squashing $\sigma^2_{z_+}/\bar{\sigma}^2_{z_0}$ by a factor of $0.57$ and anti-squashing $\sigma^2_{z_-}/\bar{\sigma}^2_{z_0}$ by a factor of $4.05$. The parametric squeezing gain $r$ can be obtained from
\begin{equation}
    \frac{\sigma^2_{z_+}}{\bar{\sigma}^2_{z_0}}=\frac{1}{1+r}\hspace{2 cm} \frac{\sigma^2_{z_-}}{\bar{\sigma}^2_{z_0}}=\frac{1}{1-r},
\end{equation}
which yields $r\approx 0.75$. This is in agreement with the ratio $r_\mathrm{max}=g/\gamma\approx 0.76$ of the parametric drive amplitude, given by the coupling rate $g/2$, to the mechanical damping $\gamma$. From the measurement, we calculate the squashing magnitude of approximately $-2.4$ dB.

\subsection{Measurement of the interparticle distance}

\begin{figure}[t]
    \includegraphics[width=\linewidth]{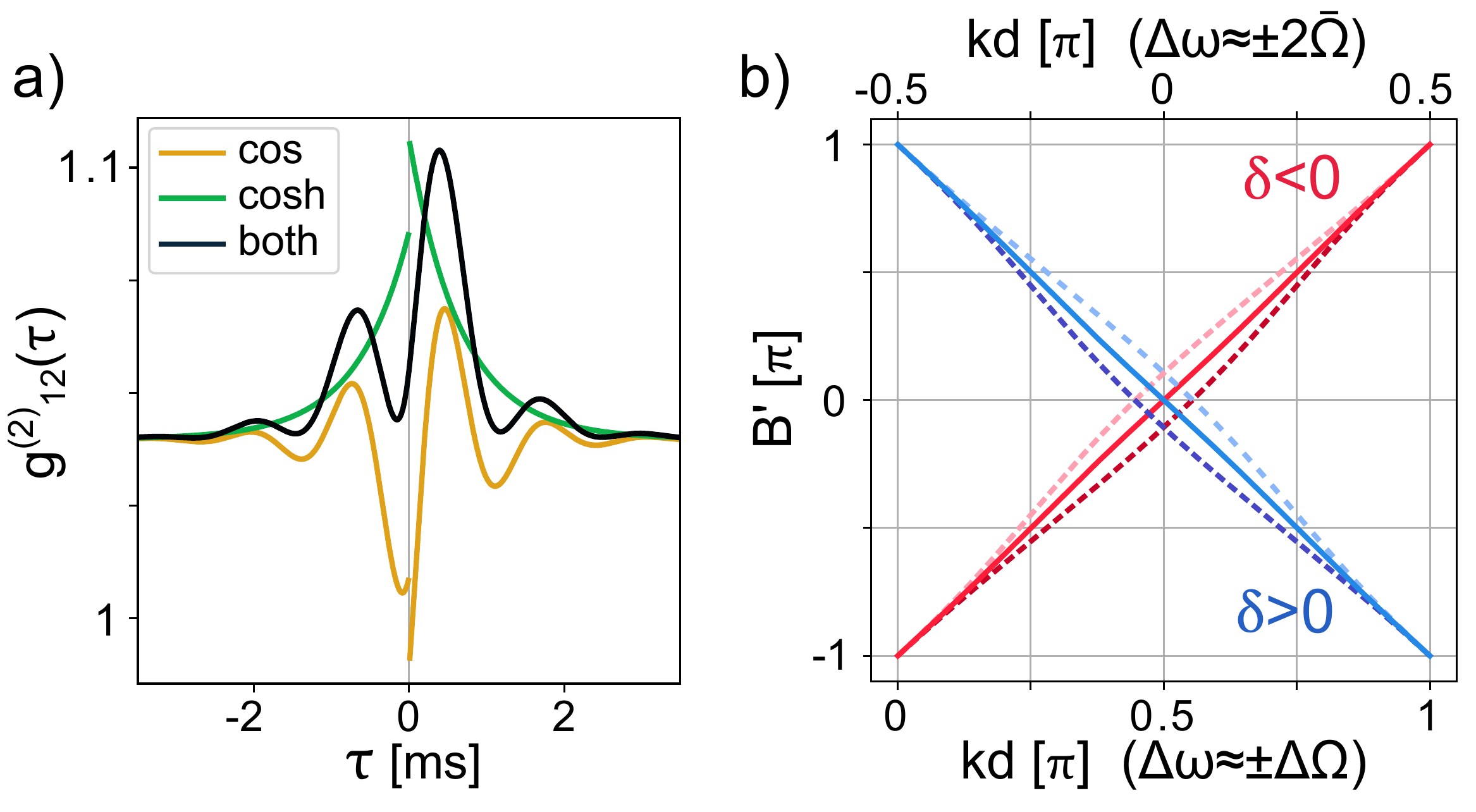}
    \caption{\textbf{Second-order cross-correlations.}
    a) The yellow and green lines show the contributions of the $\cos$ and $\cosh$ terms, respectively, to the total second-order cross-correlation function (black line) for $(\Delta\omega-\Delta\Omega)/2\pi=0.74$ kHz, $kd=0.12\pi$, $\Delta\phi=0.15\pi$, and $g/2\pi=\gamma/2\pi=0.25$ kHz. The constant offset stems from the product of particle variances.
    b) Phase offsets of the $\cos$ term as a function of $kd$ at the effective detuning of $(\Delta\omega-\Delta\Omega)/2\pi=1.51$ kHz. Regions close to the difference frequency detuning ($\Delta\omega\approx\Delta\Omega$, bottom axis) and the sum frequency detuning ($\Delta\omega\approx2\bar\Omega$, top axis) are shown. The dashed lines show $B'$ for positive (light red) and negative (dark red) times, while the solid line shows their average $\bar B'$, which is a linear function of $kd$.}
    \label{figSM:g2_th}
\end{figure}

\begin{figure}[t]
    \includegraphics[width=\linewidth]{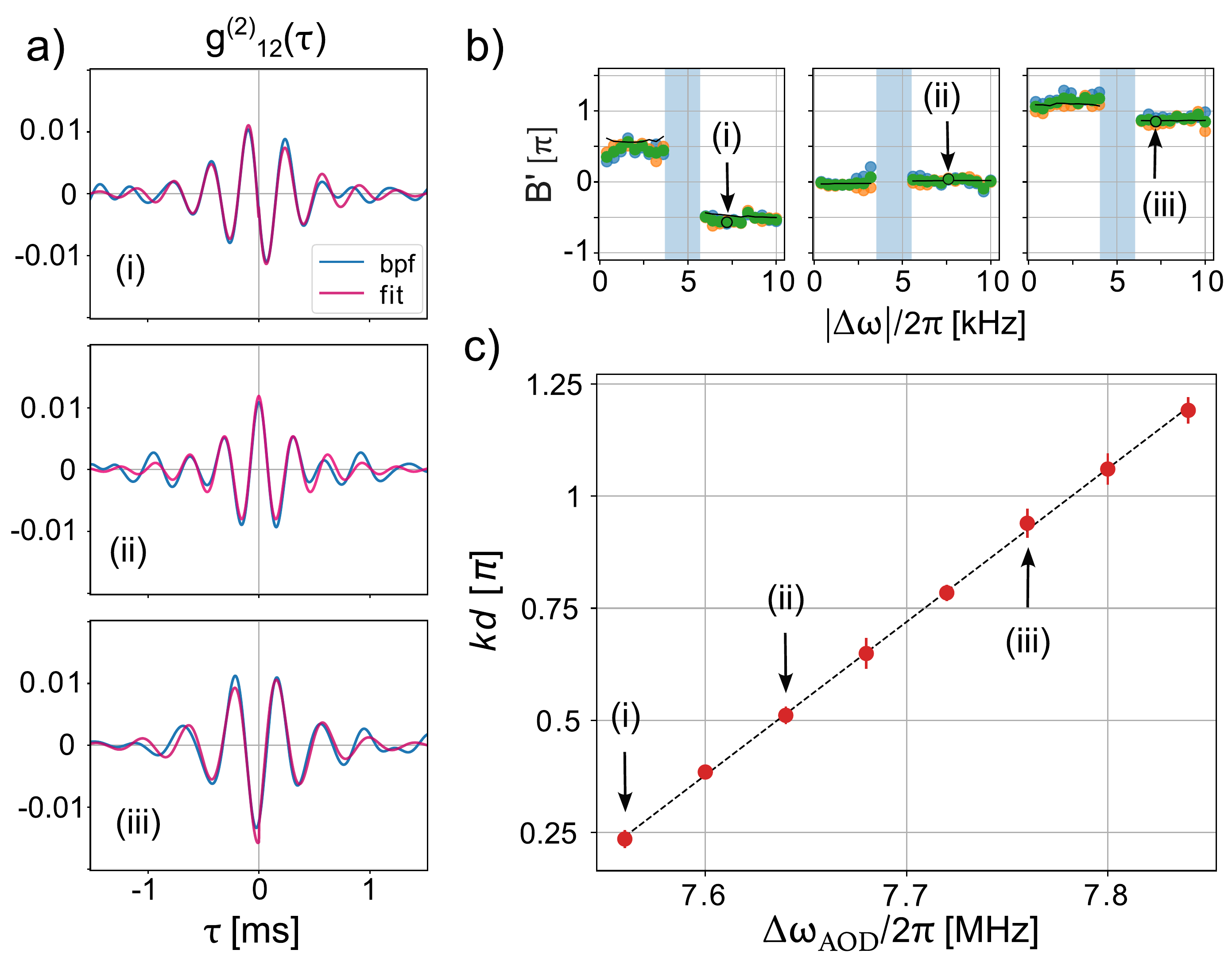}
    \caption{\textbf{Measurement of $kd$.} a) Experimental filtered second-order cross-correlations (blue line), corresponding to an optical detuning of $\Delta\omega=-7.2$ kHz at different interparticle distances, fit with the theoretical dependence (red line). b) Phases at the origin of the second-order cross-correlations $\bar B'$ as a function of the optical detuning for the three different distances shown in a). Shaded areas indicate the on-resonant excluded range where the difference between the optical and mechanical detunings is less than $1$ kHz in absolute value (mechanical detunings of $-4.7\pm0.1$ kHz, $-4.5\pm0.1$ kHz, and $-5.0\pm0.1$ kHz from left to right of panel b), respectively. The blue (orange) points correspond to the fit results for $B'(\tau<0)$ and $B'(\tau\geq 0)$, respectively; the green points are their average. c) The traveling phase due to the particle distance (modulo $2\pi$), as a function of the AOD tone frequency difference.}
    \label{figSM:g2_exp}
\end{figure}

We use the second-order cross-correlation function 
\begin{eqnarray}
    & g_{jj'}^{(2)}(\tau)=\langle |b_j(t)|^2|b_{j'\neq j}(t+\tau)|^2\rangle
\end{eqnarray}
to determine the experimental value of $kd$ and, therefore, the interaction type. We use the Isserlis-Wick theorem 
\begin{align}
    g_{jj'}^{(2)}(\tau) &= \langle |b_j(t)|^2\rangle
    \langle|b_{j'}(t+\tau)|^2\rangle+|\langle b^*_j(t)b_{j'}(t+\tau)\rangle|^2
    \nonumber \\
    &=g_{jj}^{(1)}(0)g_{j'j'}^{(1)}(0)+|g_{jj'}^{(1)}(\tau)|^2.
    \label{eq:Siegert}
\end{align}
under the assumption of Gaussian statistics of thermal noise in our system. The time-dependent part of the $g^{(2)}_{jj'}(\tau)$ is:
\begin{align}
    &|g^{(1)}_{12}(\tau)|^2 = 
    \nonumber \\
    &   |A|^2 e^{-\gamma \tau}  \left[ \cos\left(\Lambda'\tau-2B'\right)+\cosh\left(\Lambda''|\tau|-2B''\right)\right],
    \label{eq:g2_simpl}
\end{align}
where $B'=\Re(B)$, $B''=\Im(B)$, $\Lambda'$ and $\Lambda''$ can be defined either for $\Lambda_\mathrm{det}$ or $\Lambda_\mathrm{sum}$, and these equations hold for either positive or negative times.
Eq. \eqref{eq:g2_simpl} shows that the second-order cross-correlation consists of damped oscillations $\propto\cos(\ldots)$  on top of an envelope with modified exponential damping $e^{-\gamma|\tau|}\cosh(\ldots)$. Both terms are discontinuous around $\tau=0$ for $kd\neq n\pi$; however, $g^{(2)}_{jj'}(\tau)$ is continuous around $\tau=0$, as can be seen in Fig. \ref{figSM:g2_th}a). The mean of $B'$ for positive and negative times, $\bar{B}'$, depends linearly on $kd$ (Fig. \ref{figSM:g2_th}b) as
\begin{align}
    & \bar B' = \begin{dcases*}
    -\textrm{sgn}(\Delta\omega-\Delta\Omega)(2kd- \pi) \text{, } \Delta\omega\approx \Delta\Omega,
    \label{eq:B_vs_kd_dfd}
    \\
    -\textrm{sgn}(\Delta\omega-2\bar{\Omega})2kd \text{, } \Delta\omega\approx 2\bar\Omega.
    \end{dcases*}
\end{align}
Far from any considered resonance ($\delta>g$), the oscillation frequency becomes much larger than the modification of the damping ($\Re(\Lambda) \gg \Im(\Lambda) $), which allows for the separation of the cosine term by spectrally filtering $g^{(2)}_{jj'}(\tau)$ around $\Delta\omega$.
This is how we experimentally measure the second-order cross-correlation phase $B'$. The determination of $kd$ from experimental data is shown in Fig. \ref{figSM:g2_exp}.

Fig. \ref{figSM:g2_exp}a) shows experimental filtered second-order cross-correlations measured at the optical detuning $\Delta\omega=7.2$ kHz for three increasing distances between the particles (controlled via the AOD tone difference of $7.56$, $7.64$, and $7.76$ MHz from top to bottom). The signals (blue data) were obtained by calculating
\begin{equation}
    g^{(2)}_{12,\textrm{exp}}(\tau)
    =\frac{ \langle |a_1(t)|^2|a_2(t+\tau)|^2\rangle }{
    \langle |a_1(t)|^2\rangle \langle|a_2(t)|^2\rangle
    }-1,
\end{equation}
where the averaging has been performed over the whole time trace and then band-pass filtered around the effective detuning $|\delta_\mathrm{det}|=|\Delta\omega-\Delta\Omega|$ in the range of $[\sqrt{0.4\delta_\mathrm{det}^2+0.1^2},\sqrt{1.5\delta_\mathrm{det}^2+0.5^2}]$. The red lines correspond to the fit of the cosine part of Eq. \eqref{eq:g2_simpl}.
The $g_{12}^{(2)}$ function exhibits distance-dependent phases at $\tau=0$: (i) close to $\pi/2$ for $kd\approx 0.25\pi$, (ii) approximately zero for $kd\approx 0.5\pi$, and (iii) around $\pi$ for $kd\approx \pi$. The fitted phases obtained for various optical detunings at least $1$ kHz away from the mechanical detuning are shown in Fig. \ref{figSM:g2_exp}b). Blue, orange, and green points correspond to $B'$ at negative and positive times and to its mean, respectively.
The blue shaded areas mark excluded detunings, where the spectral separation of different terms in $B'$ is impossible.
The solid black line is the theoretical phase obtained for the $kd$ calculated from the mean $\bar B'$ for $\Delta\omega>\Delta\Omega$ using Eq. \eqref{eq:B_vs_kd_dfd}. 
Finally, performing this analysis for all interparticle distances and using Eq. \eqref{eq:B_vs_kd_dfd}, we obtain $kd$ as a function of the AOD tone difference, shown in Fig. \ref{figSM:g2_exp}c). These values were used in Fig. \ref{fig4:arbitrary} of the main text.

\subsection{Complex eigenfrequencies and phase correlations}

\begin{figure}[t!]
    \includegraphics[width=\linewidth]{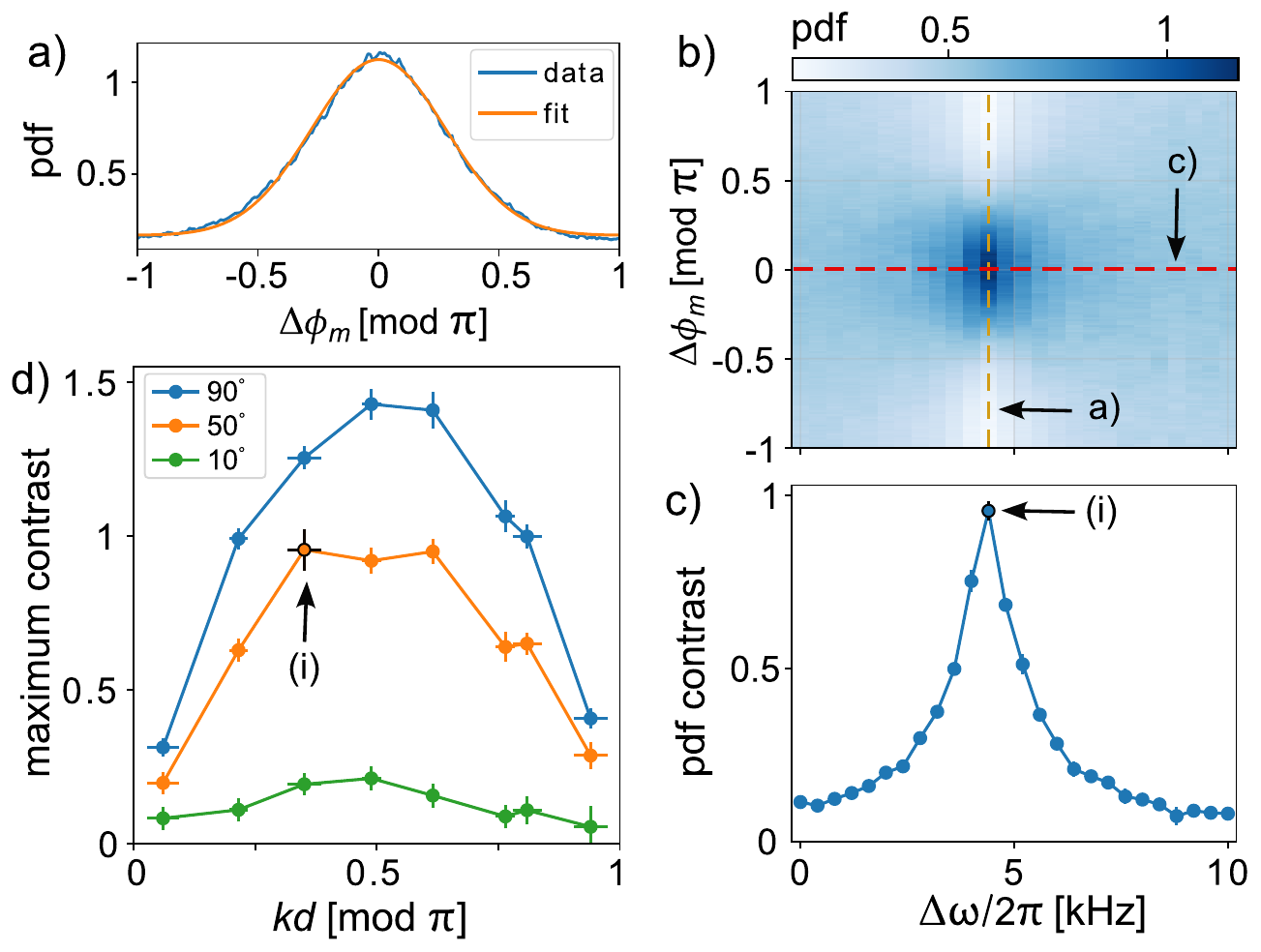}
    \caption{\textbf{Phase locking and measurement of $\Delta\phi$.}
    a) Histogram of the mechanical phase difference for $\Delta\omega_{\mathrm{AOD}}/2\pi=7.68$ MHz and $\Delta\omega/2\pi=4.4$ kHz.
    b) Optical detuning scan of the centered histograms of the mechanical phase difference for $\Delta\omega_{\mathrm{AOD}}/2\pi=7.68$ MHz and $\theta=50^\circ$.
    c) Contrast of the mechanical phase difference histograms as a function of the optical detuning.
    d) Maximal contrast from c) as a function of distance for different interaction strengths, controlled via the polarization angle $\theta$.}
    \label{figSM:phs_hst}
\end{figure}

Dissipative coupling results in modified damping rates and phase locking between the particles' motions \cite{Reisenbauer2024}. Here, we observe and quantify it across various strengths of nonreciprocal interactions and use it to determine the phase difference $\Delta\phi$ between the particles at the beginning of the measurement, which is necessary to reconstruct the eigenmodes and the corresponding squashed and anti-squashed quadratures. Note that the time-dependent interactions considered in this work make the locking phase between the bare modes $a_j$ oscillate in time. Therefore, it is necessary to switch to a rotating frame to observe stationary phase locking of the $b_j$ modes.
For measuring the phase locking, we obtain the probability histogram $P(\Delta\phi_m)$ of the instantaneous mechanical phase difference
\begin{equation}
    \Delta\phi_m(t)=\arg\left[ b^*_1(t)b_2(t) \right].
\end{equation} 
In the presence of phase locking, $P(\Delta\phi_m)$ peaks around $\Delta\phi$, which we determine by fitting the histogram with a $2\pi$-periodic Gaussian function (Fig. \ref{figSM:phs_hst}a).
We remove the phase offset from the particles' dynamics by delaying the second particle by $-\Delta\phi$.

Fig. \ref{figSM:phs_hst}b) shows a color plot of the centered probability density function $P(\Delta\phi_m)$ for various optical detunings from $0$ to $10$ kHz for a distance of $kd= (0.35\pm0.04)\pi$.

The phase-locking strength is represented by color contrast.
Its variation around the interaction resonance is shown in Fig. \ref{figSM:phs_hst}c) by plotting the contrast of the fitted PDFs, defined as 
\begin{equation}
\underset{\Delta\phi_m}{\max}(P^{\mathrm{fit}}(\Delta\phi_m,\Delta\omega))-\underset{\Delta\phi_m}{\min}(P^{\mathrm{fit}}(\Delta\phi_m,\Delta\omega))    
\end{equation}
as a function of the optical detuning.
The PDF contrast is most pronounced for $\Delta\omega/2\pi=4.4$ kHz, corresponding to the closest measured optical detuning around the mechanical detuning $\Delta\Omega=4.5\pm0.1$ kHz.
In Fig.\ref{figSM:phs_hst}d), we assessed the resonant phase locking strength for different distances by plotting the maximum contrast over the entire optical detuning range as a function of the reconstructed distance phase $kd$. Different colors correspond to different interaction strengths $g$ controlled by the laser polarization angle $\theta$. The larger error bars than in Fig.\ref{figSM:phs_hst}c) account for possible deviations from optimal optical detuning due to the detuning step size of $400$ Hz.
The phase-locking strength qualitatively follows the magnitude of the imaginary part of the eigenfrequency; it is maximal around $kd=\pi/2$, and increases with the interaction strength.

\subsection{Detection and cooling}

We detect the particles' motion by collecting the trapping light reflected off the particles and then separating it from the incoming laser beams using a Faraday rotator. To enable independent detection of each particle's motion, we first re-image the trapping plane onto a right-angled prism that separates the scattered light beams into two directions. We subsequently re-image the prism plane into two single-mode fibers, which perform mode cleaning for each beam \cite{Reisenbauer2024}. The collected light is mixed with a local oscillator inside a 50:50 fiber beamsplitter to realize balanced homodyne or heterodyne detection. Heterodyne detection is used to read out the particle's motion, while phase-locked homodyne detection is used for feedback cooling. 

We probe the charge deposited on the particles by applying an electrostatic force with an electrode positioned next to our trapping lens. For that, the electrode is driven with an oscillating voltage near the particles' mechanical resonances. If a response in the motion of only one particle is observed, we conclude that the other particle is neutral. When both particles are driven, a plasma discharge is generated within the vacuum chamber, randomly altering the charge on both particles. We continuously measure the particle charges and switch off the drive when only one particle is charged.

We can reduce the charged particle's variance through feedback cooling with cold damping. The feedback signal is generated by filtering and phase-shifting the detected homodyne signal with a RedPitaya. This introduces additional dissipation into the particle's motion, resulting in an increased total damping rate $\gamma_{\text{eff}} = \gamma + \gamma_{\text{fb}}$. The time-dependent protocol presented in Fig. \ref{fig3:extremes}a is driven by a $40$ Hz square-wave signal acting as a trigger. The trigger switches the cooling on and off with a period of $25$ ms. A second trigger, phase-shifted to compensate for differences in the electronic paths, controls the frequency of one RF tone to instantaneously turn off the interaction by moving $\Delta\omega$ far away from any resonance.

To describe the non-equilibrium evolution of the particle variances ($\tau=0$), the time $t$ in the first-order correlation function will account for the time elapsed after switching on the interactions and stopping the cooling of one particle. We calculate the first-order correlation functions by averaging over many repetitions, where the statistical properties of the initial state ($ \langle b^*_{j}(0)b_{j'}(0) \rangle =\delta_{jl} n^-_{j}$) do not vanish as $t$ remains a relevant variable due to time-dependent evolution. For $t<0$, the cooled particle has achieved an occupation $n^-_{j}(t<0)=n_j\gamma/\gamma_j$, where $\gamma_j=\gamma_\mathrm{fb}+\gamma$ is the total damping rate due to feedback cooling and gas damping. For $t>0$, the damping is equal for both particles. We obtain
\begin{align}
 &\langle|b_j(t)|^2\rangle 
    =
    \Big(n^-_j|\alpha_j(t)|^2+n^-_{j'} |\beta_j(t)|^2\Big)e^{-\gamma t}+ \nonumber \\
    & \gamma\int\limits_{0}^t dt' e^{-\gamma t'}\left[n_j|\alpha_j( t')|^2+n_{j'}|\beta_j( t')|^2
    \right]
    \label{eq:quench11}
    \\
 &\langle b^*_1(t)b_2(t)\rangle=\Big(n^-_1\alpha^*_1(t)\beta_2(t)+n^-_2\beta_1^*(t)\alpha_2(t)
    \Big)e^{-\gamma t}+ \nonumber \\
    & \gamma \int\limits_{0}^t dt' e^{-\gamma  t'}\left[n_1\alpha^*_1( t')\beta_2( t')+n_2
    \beta_1^*( t')\alpha_2( t')\right],
    \label{eq:quench12}
\end{align}
where the first terms account for the initial conditions, while the second terms account for the final stationary variance. We fit the experimental data in Fig. \ref{fig3:extremes}a) in the main text with functions $\langle|b_j(t)|^2\rangle$.

\end{document}